\documentclass[preprint]{aastex}
\newcommand{\wcen}{$\omega$~Cen~}
\newcommand{\mga}{$^{24}$Mg}
\newcommand{\mgb}{$^{25}$Mg}
\newcommand{\mgc}{$^{26}$Mg}

\usepackage[dvips]{color}

\shorttitle{MAGNESIUM ISOTOPE RATIOS IN $\omega$ CENTAURI RED GIANTS}

\shortauthors{Da Costa, Yong \& Norris}

\begin{document}

\title{MAGNESIUM ISOTOPE RATIOS IN $\omega$ CENTAURI RED GIANTS}

\author{G. S. Da Costa, John E. Norris and David Yong}

\affil{Research School of Astronomy and Astrophysics, Australian 
National 
University, Canberra, ACT 0200,
Australia}

\slugcomment{Accepted for publication in The Astrophysical Journal}

\begin{abstract}

We have used high resolution observations obtained at the AAT with UHRF ($R$ $\sim$ 100,000) 
and at Gemini-S with b-HROS ($R$ $\sim$ 150,000) to determine magnesium isotope ratios for seven
\wcen red giants that cover a range in iron abundance from [Fe/H] = --1.78 to --0.78 dex, and for
two red giants in M4 (NGC~6121).  The \wcen stars sample both 
the ``primordial'' (i.e., O-rich, Na and Al-poor) and the ``extreme'' (O-depleted, Na and Al-rich) populations
in the cluster.  The primordial population stars in both \wcen and M4 show (\mgb, \mgc)/\mga\/ isotopic 
ratios that are consistent with those found for the primordial population in other globular clusters with
similar [Fe/H] values.  The isotopic ratios for the \wcen extreme stars are also consistent 
with those for extreme population stars in other clusters.  The results for the extreme 
population stars studied indicate that the \mgc/\mga\/ ratio 
is highest at intermediate metallicities ([Fe/H] $<$ \mbox{--1.4} dex), and for the highest
[Al/Fe] values.   Further, the relative abundance of \mgc\/ in the extreme population stars is notably higher
than that of \mgb, in contrast to model predictions.  The \mgb/\mga\/ isotopic ratio in fact does not show 
any obvious dependence on either [Fe/H] or [Al/Fe] nor, intriguingly, any obvious difference between 
the primordial and extreme population stars.

\end{abstract}


\section{INTRODUCTION}

The stellar system $\omega$ Centauri has been known to have an unusual stellar population
 for almost four decades.  For example, there is now an extensive body of work which shows
that, unlike the situation for most globular clusters, the member stars of \wcen possess 
a large range in heavy element abundance together with distinctive element-to-iron abundance
ratios \citep[e.g.,][and the references therein]{JP10,AM12}.  Based on photometric and spectroscopic
studies of lower main sequence stars there is also evidence for a substantial spread in helium
abundance in the cluster \citep[e.g.,][]{JEN04, GP05}.  Further, spectroscopic studies of stars near the
main sequence turnoff suggest that \wcen has an age spread of order 2 Gyr \citep[e.g.,][and the
references therein]{EP11}.
These properties, which are not shared, in toto, by any other globular cluster, have led to
the suggestion that \wcen is the nuclear remnant of a disrupted dwarf galaxy that was accreted by
the Milky Way \citep[e.g.,][]{KCF93}.  \citet{BF03} have shown that this is dynamically plausible 
despite the tightly bound and retrograde current orbit of the cluster \citep{Di99}.

In one respect, however, \wcen is very similar to the other globular clusters of the Milky Way
and that is the presence of the so-called O-Na anti-correlation among the cluster stars
\citep{ND95a}.  It is now well-established \citep[e.g.,][]{CB10} that essentially all Milky Way 
globular clusters
show this phenomenon.  It is characterized by the presence of two or three populations within a given
globular cluster: {\it primordial}, {\it intermediate} and, in at least some clusters, an
{\it extreme} population \citep[e.g.,][]{CB09}.  The primordial population shows abundances of the elements 
C, N, O, F, Na, Al and Mg relative to Fe that are essentially equivalent to those for field halo stars with the 
same [Fe/H] value.
The intermediate population shows depletions in O of order 0.3 dex, and depletions in C and F, which are 
coupled with enhancements in N, Na and Al.  The extreme population shows even larger depletions in O 
and also depletions in the Mg abundance in addition to the N, Na and Al enhancements. 

The usual interpretation of these abundance correlations is that they are the consequence of
proton-capture reactions occurring near the H-burning shell in some astrophysical source (as originally
suggested in this context by \citet{DD90} and \citet{LHS93}).  The reactions
involved are those of the Ne-Na and Mg-Al cycles \citep[e.g.,][]{Ca96}.
Magnesium abundances, and in 
particular Mg isotope ratios, have played a crucial role in the development of theories for the origin of the
abundance anomalies.  In particular, the outcome of 
the pioneering work of \citet{MS96} helped rule out individual star ``deep mixing'' scenarios as a possible
explanation for the observed abundance differences.  
Specifically, \citet{MS96} analysed spectra of two O-rich and
four extremely O-poor red giants in M13.  The O-poor stars were also strongly Al-enhanced and notably
Mg-depleted.  The key result was the recognition that the Mg-depletions in the O-poor stars required 
depletions in the \mga\/ abundance, with the \mgb\/ and \mgc\/ isotopes having above solar abundance
ratios.  Since
the {\it p}-capture reaction on \mga\/ in the Mg-Al cycle is strongly temperature dependent, significant
\mga\/ depletions occur only for temperatures well above those found near the H-burning shell in 
canonical low mass stellar models.  Consequently, while evolutionary mixing does occur in red giant 
branch stars, the primary origin of the abundance anomalies is now 
accepted as the result of events occurring during the formation of globular clusters that involve the 
inclusion of material processed in either rapidly rotating massive stars \citep[e.g.,][]{De07}, massive 
binary stars \citep{dM09} or relatively massive AGB stars \citep[e.g.,][]{VDA05} into a second generation
of cluster stars.  Subsequent work by \citet{DY03, DY06} on the Mg isotope
ratios for red giants in the clusters NGC~6752 and M13, and by \citet{MC09} for M71, have 
verified and extended the results of \citet{MS96}.

While there are extensive studies of the O-Na anti-correlation in a number of clusters with different [Fe/H]
values \citep[e.g.,][]{CB09}, the presence of a [Fe/H] abundance range in $\omega$~Cen can be turned
to advantage as it allows investigation of the O-Na anti-correlation phenomenon as a function of [Fe/H] 
in a single 
(admittedly complex) system.  The richness of $\omega$~Cen also allows large samples of stars to be
studied.  The investigations of \citet{JP10} and \citet{AM11} show that the O-Na correlation is present at 
all [Fe/H] values except perhaps for the highest [Fe/H] values ([Fe/H] $\gtrsim$ --1.0 dex).  In particular, the
maximum  value of [Na/Fe] seen in O-depleted stars increases monotonically with [Fe/H] 
reaching [Na/Fe] $\sim$ +1.0 for the most 
metal-rich objects.  In this latter population Na-rich stars are dominant despite the apparent presence of
a significant range in [O/Fe] values.  The situation for [Al/Fe] is somewhat different -- the maximum value
of [Al/Fe] in O-poor, Na-rich stars is relatively constant at [Al/Fe] $\sim$ +1.2 for stars with [Fe/H] 
$\lesssim$ --1.3, but for higher metallicities the maximum value of [Al/Fe] decreases significantly to 
[Al/Fe] $\sim$ +0.6 at the highest metallicities \citep{ND95b, JP10, DM11}.
Possible ways to generate these abundance patterns are discussed in, for example, \cite{JP10} and 
\citet{DA11}. 

The only extensive study of Mg in $\omega$~Cen stars, however, is that of \citet{ND95b}\footnote{We 
note for completeness that \citet{VS00} derived Mg abundances for 10 $\omega$~Cen
red giants while \citet{BW93} provided Mg abundances for 6 stars, one in common with \citet{VS00}.}.  
The results for the stars analysed by these authors 
(whose data for Na, Al and O exhibit the same trends with [Fe/H] as those of \citet{JP10} and \citet{AM11}) 
reveal only a minor population of stars with significant Mg depletions.  
The stars with Mg-depletions are, as
expected, strongly enhanced in [Al/Fe] and generally also show significant depletions in [O/Fe].  
In the terminology of \citet{CB09} most belong to the extreme population.  While the sample is not large
compared to those of \citet{JP10} and \citet{AM11}, it is notable that these Mg-depleted stars are all 
relatively metal-poor: the highest [Fe/H] value is --1.4 dex.  This is no doubt related to the smaller Al
enhancements seen at higher [Fe/H] values \citep{ND95b, JP10, DM11}.

It is clear that further information on the process or processes underlying the O-Na anti-correlation could
be obtained by the measurement of Mg isotope ratios in a suitably selected set of $\omega$~Cen red 
giants.  For example, in the 6 M$_{\sun}$ metal-poor AGB model of \cite{V11}, Mg-Al cycle burning at the 
bottom of the convective envelope produces surface Mg abundances that are dominated by \mgb\/ with
low \mgc\/ abundances.  This is in conflict with the existing data for Al-rich stars in NGC~6752 and
M13 where \mgc\/ has a higher abundance than \mgb\/ \citep{DY03, DY06}.  Confirmation of this 
discrepancy would likely drive a reassessment of the theoretical treatments, such as mixing
processes and nuclear reaction rates, employed in the stellar models.

Against this background, the purpose of the present paper is to determine the Mg isotope ratios for a 
representative set of red giants in $\omega$~Cen and for two red giants in the ``normal'' globular cluster
M4 (NGC~6121).  The selection of the sample is described in Section 2, 
together with a presentation of the observations and data reduction.  The analysis to derive the Mg
isotope ratios is discussed in Section 3 and the results for the $\omega$~Cen and M4 stars are presented in Section 4.   In Section 5 we combine our results with previous studies and discuss the implications 
in the context of existing models for the synthesis of the \mgb\/ and \mgc\/ isotopes.

\section{OBSERVATIONS AND REDUCTIONS}

\subsection{Sample Selection}

The $\omega$~Cen stars to be observed in this program were selected from the set of red giant members
observed at high resolution and analysed by \citet{ND95b}.  These authors give atmospheric parameters, 
iron abundances [Fe/H], and element-to-iron abundance ratios for a biased sample of 40 stars that 
covers nearly the full range of [Fe/H] values seen in the cluster.   In particular, we first selected ROA~94, 
which with [Fe/H] = --1.78 is one of the most metal-poor stars in $\omega$~Cen \citep{ND95b,JP10,AM12}.  
The star has an enhanced 
 [$\alpha$/Fe] ratio and low values for [Na/Fe], [Al/Fe] and [Ba/Fe] \citep{ND95b}.  It is therefore a member 
 of the {\it primordial} population of the cluster in the terminology of \citet{CB09}, and with the low [Fe/H]
 value, a good 
 candidate with which to characterise the ``initial'' value of the Mg isotope ratios in the cluster.  The 
 remaining stars in the observing list were pairs of stars selected to have similar [Fe/H] values but with
 significantly different [O/Fe], [Na/Fe], etc, abundances.  In other words, we selected pairs of {\it primordial} 
 and {\it extreme} population stars with similar [Fe/H] values to investigate the difference in the Mg isotopic ratios 
 between the different populations.  The set of pairs also encompassed a range of [Fe/H] values so that any
 difference between the populations as a function of [Fe/H] could be explored.  Three such pairs of stars 
 were subsequently observed: ROA~43 and 100, ROA~132 and 150, and ROA~201 and
 248.  Their abundance characteristics and atmospheric 
 parameters along with those for ROA~94 are given in Table \ref{Table 1}.  The values are taken from \citet{ND95b}.  
 
 We note that the primordial stars ROA~43, ROA~132 and ROA~201 belong to the 
``CO-strong'' population in $\omega$~Cen, while the extreme stars ROA~150 and ROA~248 belong to the
``CO-weak'' population.  No CO-strength classification is available for ROA~94 or ROA~100.  As discussed
in \citet{PC80} and \citet[][Section 2.1]{ND95b}, the $\omega$~Cen CO-strong stars have stronger 
near-IR CO indices than those of all of the red giants in other globular clusters for which \citet{PC80} 
present data, while the CO-weak stars have CO indices comparable to those of other globular cluster red giants at similar 
[Fe/H] values.  For completeness we note also that the extreme stars ROA~100 and ROA~150 are 
classified ``CN-strong'' in \citet{ND95b} while the primordial star ROA~94 is classified ``CN-weak''.  No
CN-band strength classification information is available for the primordial stars ROA~43, ROA~132 and 
ROA~201 and for the extreme star ROA~248.

Six of our seven stars have also been observed in \citet{JP10}: stars ROA~94, 43, 100, 132, 150 and 248
correspond to \citet{JP10} stars 39037, 10012, 44449, 55114, 73025 and 55071, respectively.  For these six stars we find mean [Fe/H], [O/Fe] and [Na/Fe] differences, in the sense \citet{ND95b} -- \citet{JP10} of 
0.00, --0.01 and +0.03 dex, respectively.  The standard deviations are 0.10, 0.13 and 0.09 dex, which are consistent with the combined uncertainties in the abundance determinations.  Although \citet{JP10} do
not list [Al/Fe] ratios for the six common stars, [Al/Fe] values are given in \citet{JP08}.  Using those data
we find a mean difference in [Al/Fe] in the sense \citet{ND95b} -- \citet{JP08} of --0.08 dex though with a
disturbingly large standard deviation of 0.47 dex.  Since the [Al/Fe] determinations of \citet{ND95b} are
in good agreement with those in \citet{VS00} (mean difference of --0.07 dex with $\sigma$ = 0.16 dex
for 3 stars which cover an [Al/Fe] range of $\sim$1 dex), we ascribe the large $\sigma$ primarily to
large uncertainties in the \citet{JP08} [Al/Fe] determinations.  In that context it is worth noting that the
\citet{JP08} spectra have only moderate spectral resolution $R$ $\sim$ 13,000 as distinct from 
$R$ $\sim$ 18,000 for \citet{JP10} and $R$ $\sim$ 38,000 for \citet{ND95b}.
    Similarly, star ROA~132 is the only one from our sample in common with the work of \citet{AM11}; it corresponds to their star 206284.  
The abundance differences in the sense \citet{ND95b} 
-- \citet{AM11} are 0.19, --0.18 and +0.07 dex for [Fe/H], [O/Fe] and [Na/Fe], respectively.  Unfortunately,
this star is not among those with C, N and O abundance measures in \citet{AM12}.

The two M4 stars were selected as the brightest pair of ``CN-weak'' (L1514) and ``CN-strong'' (L2406)
stars in the detailed abundance study of M4 red giants carried out by \citet{II99}.  Based on the 
[O/Fe] and [Na/Fe] values given in \citet{II99}, star L1514 is a member of the primordial population of M4
while L2406, which has an oxygen abundance 0.22 dex lower than L1514 and a sodium
abundance 0.30 dex higher, is a member of the cluster's intermediate population.  The two stars
differ marginally in [Al/Fe] with L2614 0.11 dex higher, but have essentially identical Mg abundances.  
Indeed in the sample of M4 red giants observed by \citet{II99}, there is no evidence for any intrinsic
range in [Mg/Fe] or any obvious correlation between [Mg/Fe] and [Na/Fe] or [Al/Fe] \citep{II99}. 
The atmospheric parameters and abundances for the two stars from \citet{II99} are given in Table
\ref{Table 1}.  The two M4 stars were also included in the study of \citet{DY08} who found similar results
to \citet{II99}.  In particular, \citet{DY08} find an oxygen abundance difference of 0.18  dex  (in the sense 
L1514 -- L2406) and [Na/Fe], [Mg/Fe] and [Al/Fe] differences of --0.36, --0.08 and --0.03 dex, respectively.
The oxygen and sodium abundances confirm the population classification for these stars.

Star L1514 is also included in the M4 study of \citet{AM08}, but L2406 is not.  Using the 
oxygen and sodium abundances of \citet{AM08}, L1514 is again classified as a primordial star.  
As did \citet{II99} and \citet{DY08}, \citet{AM08} find no significant magnesium - aluminium 
anti-correlation, and that [Mg/Fe] does not correlate with either [O/Fe] or (anti)correlate with [Na/Fe] 
within their extensive sample of M4 red giant members.  
The total [Al/Fe] range is also small at $\lesssim$0.4 dex \citep{AM08}.
 
\begin{deluxetable}{lcccrrrrrc}
%\rotate
\tablewidth{0pt}
\tablecaption{Globular Cluster Program Stars \label{Table 1}}
\tablecolumns{10}
\tablehead{
\colhead{Object} & \colhead{T$_{eff}$} & \colhead{log g} & \colhead{[Fe/H]} & \colhead{[C/Fe]} 
& \colhead{[O/Fe]} & \colhead{[Na/Fe]} & \colhead{[Mg/Fe]} & \colhead{[Al/Fe]}  & \colhead{Pop Type}}
%& & & \colhead{(km s$^{-1}$)} & \colhead{(km s$^{-1}$)} & & \colhead{(mag)} & \colhead{(mag)}
%& \colhead{(\AA)}  & \colhead{(\AA)}    }
\startdata
$\omega$ Cen ROA 94 & 4200 & 0.7 & --1.78 & --0.67 & 0.25 & --0.26 & 0.51 & $<$0.17 & P \\ 
 & & & & & & & & \\
$\omega$ Cen ROA 43 & 3950 & 0.4 & --1.47 & --0.21 & 0.39 & 0.21 & 0.52 & 0.06 & P \\
$\omega$ Cen ROA 100 & 4150 & 0.7 & --1.49 & $<$--0.54 & $<$--0.43 & 0.58 & 0.05 & 1.15  & E \\
 & & & & & & & & \\
$\omega$ Cen ROA 132 & 3900 & 0.3 & --1.37 & --0.22 & 0.35 & 0.17 & 0.56 & --0.32 & P \\
$\omega$ Cen ROA 150 & 3950 & 0.6 & --1.25 & --0.83 & --0.46 & 0.59 & 0.26 & 1.06 & E \\
 & & & & & & & & \\
$\omega$ Cen ROA 201 & 3750 & 0.5 & --0.85 & --0.30 & 0.37 & --0.12 & 0.45 & --0.07 & P \\
$\omega$ Cen ROA 248 & 3850 & 0.6 & --0.78 & \nodata & --0.33 & 0.75 & 0.24 & 0.68 & E \\
& & & & & & & & & \\
NGC 6121 L1514 & 3875 & 0.3 & --1.18 & --1.00 & 0.41 & 0.01 & 0.40 & 0.44 & P \\
NGC 6121 L2406 & 4100 & 0.5 & --1.18 & --0.84 & 0.19 & 0.31 & 0.37 & 0.55 & I \\
& & & & & & & & & \\
NGC 6752 mg0 & 3928 & 0.3 & --1.62 & \nodata & --0.08 & 0.65 & 0.46 & 1.06 & E \\
\enddata
\tablecomments{Except for the last  column, the data for the $\omega$~Cen entries are taken 
from \citet{ND95b}, while those for the M4 stars come from \citet{II99}.  The NGC~6752 mg0 entries
use the results of \citet{DY03}, including the solar 
abundances given in that paper.  In the last column  ``P'' indicates the star is classed as belonging to the 
primordial population, ``I'' for the intermediate population and ``E'' for stars in 
the extreme population, using the terminology of \citet{CB09}.  $\omega$~Cen stars  ROA~132 
and ROA~150 were observed with UHRF at the AAT as were the two M4 (NGC~6121) stars.
$\omega$~Cen stars ROA~94, 100, 201 and 248, together with NGC~6752 mg0, were observed with 
b-HROS at Gemini-S\@.  Star ROA~43 was observed at both facilities. }
\end{deluxetable}

\subsection{Observations}

The observations were obtained with two different telescope/spectrograph combinations.  The first 
employed the ``Ultra-High Resolution Facility'' (UHRF)\footnote{www.aao.gov.au/astro/uhrf.html} 
on the Anglo-Australian Telescope (AAT)\@.  The instrument
was used in the resolution 300,000 mode but with a 0.6$\arcsec$ slit in place of the normal imager slicer.
This setup yields spectra with an effective resolution $R$ $\sim$ 100,000 and a scale of 0.018\AA\/
per binned pixel with the Tek 1k$\times$1k detector in use at the time.  Due to the limited wavelength 
coverage available with this detector, the central wavelength of the spectrograph setup had to be
adjusted for the geocentric velocity of each target to ensure the isotopic abundance sensitive MgH
features ($\lambda$ $\sim$ 5134.55 -- 5140.2\AA) were centred.   
Data were obtained in two runs in 2000 April and 2001 March  
under seeing conditions that varied from 1.2$\arcsec$ to 2.4$\arcsec$, with a typical value of 1.8$\arcsec$.
As the small slit size compared to the seeing limits the instrument efficiency, total exposure times of 
many hours (broken into individual 1800 sec integrations) were needed to 
achieve signal-to-noise (S/N) values that would be useful for this project.  In the event, as we shall see 
below, we were able to obtain S/N values of $\sim$50--60 per pixel for the $\omega$~Cen stars
ROA~43, 132 and 150 with this instrument setup for total exposure time of 10, 10 and 17.5 hrs,
respectively.  For the M4 stars L1514 and L2406, the S/N values were $\sim$70 and $\sim$45 for
total exposure times of 2.0 and 3.5 hours, respectively.  We also obtained spectra of bright field stars for
which Mg isotope ratios have been determined by \citet{GL00}.  Flat field, Th-Ar arc lamp and dark 
frame exposures were obtained each night during both observing runs. 

The second telescope/instrument combination used was the (now decommissioned)
b-HROS\footnote{www.gemini.edu/sciops/instruments/hros/hrosindex.html, accessible via www.gemini.edu/node/10002},
bench-mounted fiber-fed high dispersion spectrograph on the Gemini-S 8m telescope.  The
spectrograph was located in the pier of the telescope and fed via fibres from a cassette mounted in
GMOS-S that locates them in the focal plane of the telescope. Guiding and tip/tilt image motion
compensation is achieved through use of the GMOS-S On-Instrument Wavefront Sensor.  We used
the `object only' mode in which a single fibre with a projected diameter of 0.9$\arcsec$ on the sky
transfers the star light to the spectrograph.  An image-slicer at the spectrograph entrance slit then 
reformats the fibre output into seven 0.14$\arcsec$ wide slices with a nominal resolution of 
$R$ $\sim$ 150,000.  The central wavelength of the 
echelle was set to $\lambda$5135\AA\/ in order 40.  With the single 2048 $\times$ 4603 pixel 
(spatial $\times$ wavelength) CCD detector this provided sufficient wavelength coverage that no 
adjustment in the central wavelength was required to compensate for the range of radial velocities 
exhibited by the various targets.  On-chip binning of $\times$4 in the spatial direction was used and
the scale in the spectral direction was 0.010\AA\/ per pixel.

The observations were carried out during a two night (2006 UT March 16 and 18) classical run.  
Both nights were clear with excellent seeing ($\sim$0.6$\arcsec$). 
Typical total exposures for the program targets were of 
order 2 hours per star, broken into individual 1800 sec integrations.  Observations were obtained of 
$\omega$~Cen red giants ROA~43, 94, 100, 201 and 248, where we note that ROA~43 was 
also observed with the AAT/UHRF set up.   In addition, we also observed the globular cluster red giant 
NGC~6752 mg0 which has been analysed by \citet{DY03}, whose data we present for comparison 
purposes in the final row of Table \ref{Table 1}.  Spectra were also obtained of three 
bright field stars for which Mg isotope ratios have been determined by \citet{GL00} or by \citet{DY03}. 
Flat field and arc lamp exposures were also obtained each night using the G-CAL facility calibration
unit.  All frames were bias-subtracted and trimmed of the overscan regions at the telescope.

\subsection{Data Reduction}

The data for the AAT/UHRF and Gemini-S/b-HROS configurations were
reduced by using the FIGARO
package\footnote{www.aao.gov.au/figaro}.  Individual
spectra were flatfielded and wavelength calibrated with spectra of the
quartz and arc lamps, respectively; they were cosmic-ray cleaned using
``bclean''; curvature in the spectra was removed with ``sdist'' and
``cdist''; individual observations of each object were wavelength
shifted to bring them into the restframe; and for each object the
individual spectra were co-added, sky-subtracted, and finally
continuum normalized by using spectra of a rapidly rotating B star
($\beta$~Centauri) to constrain the shape of the instrumental response
function.  An essential difference between the two instrumental
configurations is that for UHRF the spectral lines were set
perpendicular to the dispersion direction on the detector, while for
b-HROS this was not the case.  For the latter, we used
curvature-corrected arc spectra to determine and remove the slope of
the spectral lines in the spatial direction by estimating the
differential wavelength shifts in that direction, and shifting the
stellar spectra by the appropriate amounts before summing the counts.

For the three $\omega$~Cen stars observed at the AAT with UHRF the net counts in the continuum near 
5140\AA\/ lie in the range 2700--3100 electrons per 0.018\AA\/ pixel.  For M4~L1514 and L2406, the net counts were 4800 and 2200 electrons per pixel, respectively.  This suggests the S/N values for these AAT globular cluster red giant spectra lie in the range $\sim$50--70 per pixel.  
For the Gemini-S data the net counts range from 6300 (ROA~43) to 15900 (NGC~6752-mg0) electrons per 0.010\AA\/ pixel, suggesting S/N in the range 
$\sim$80--130 per pixel.  These values are confirmed by a direct determination of the RMS dispersion 
of the counts in
the stellar continuum in small bands of width $\sim$0.3\AA\, at
5135.5\AA\/ and 5137.9\AA\/ in the spectra of the hotter and higher S/N $\omega$~Cen stars
ROA~94 and 100 and in NGC~6752 mg0.  The RMS dispersions lead to S/N values in the
range 90--130, consistent with those calculated from the net counts.

\begin{figure}
\centering
\includegraphics[width=0.7\textwidth]{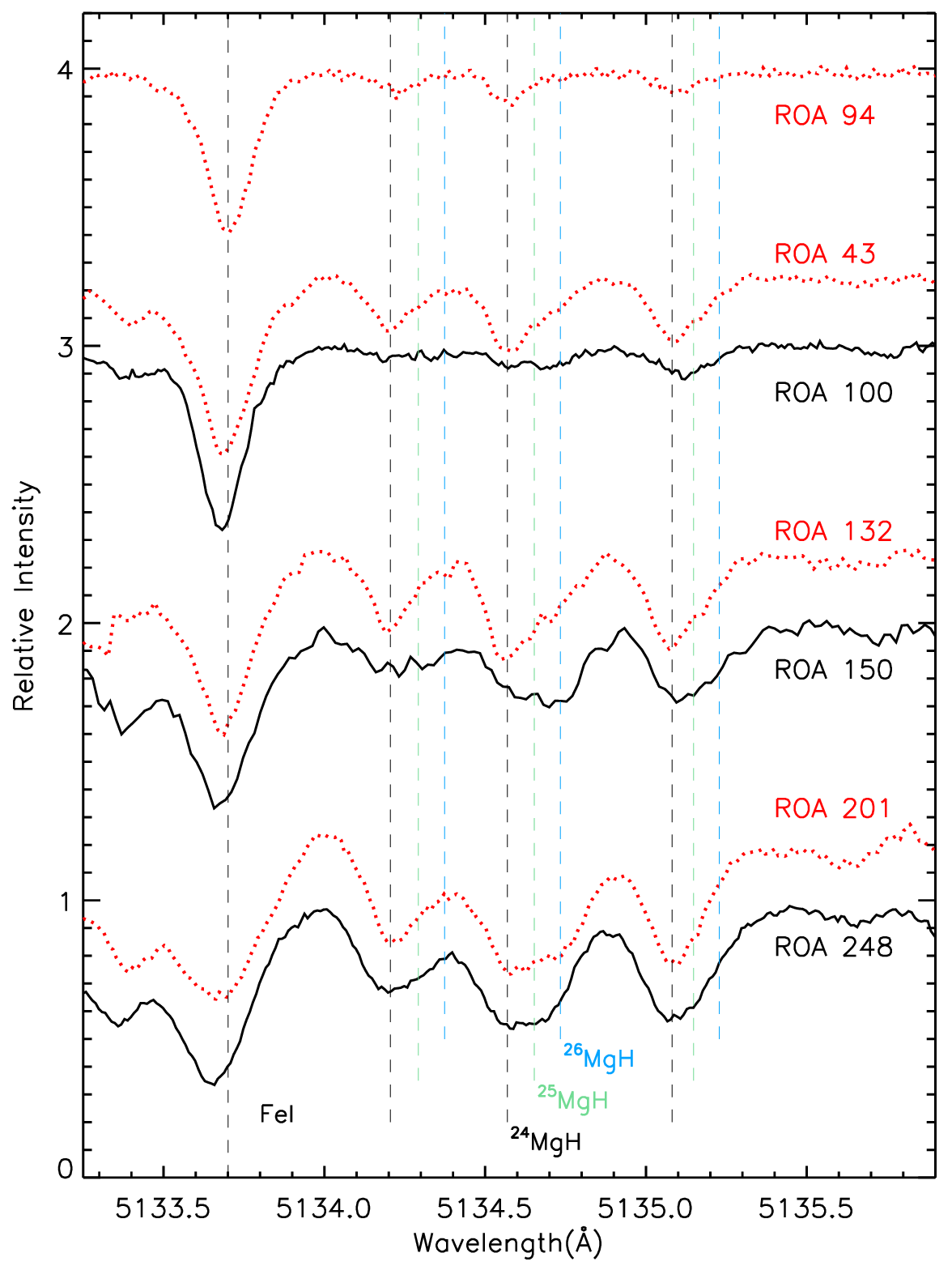}
\caption{Spectra for the $\omega$ Cen program stars for wavelengths between 5133\AA\/ 
and 5136\AA. The spectra 
are ordered approximately by metallicity with ROA 94 being the most metal-poor and 
ROA 248 the most metal-rich, respectively.  The primordial population star spectra
are plotted as red dots, while those for the extreme population stars are plotted as black lines. The
positions of the $^{24}$MgH, $^{25}$MgH, and $^{26}$MgH features are marked by
vertical lines. 
\label{DY1}
}
\end{figure}

\begin{figure}
\centering
\includegraphics[width=0.7\textwidth]{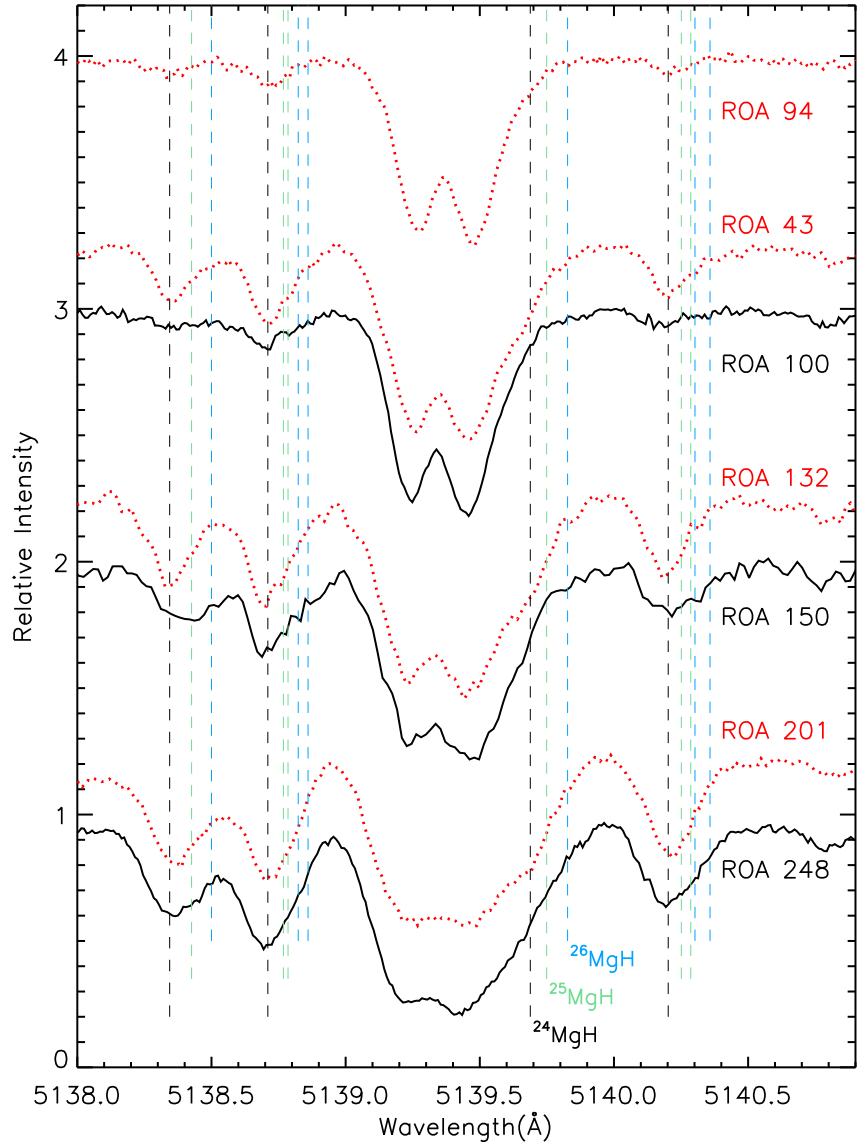}
\caption{As for Fig.\ \ref{DY1} except that the wavelength region shown is 5138\AA\/ to
5141\AA.  \label{DY2}
}
\end{figure}

\begin{figure}
\centering
\includegraphics[width=0.7\textwidth]{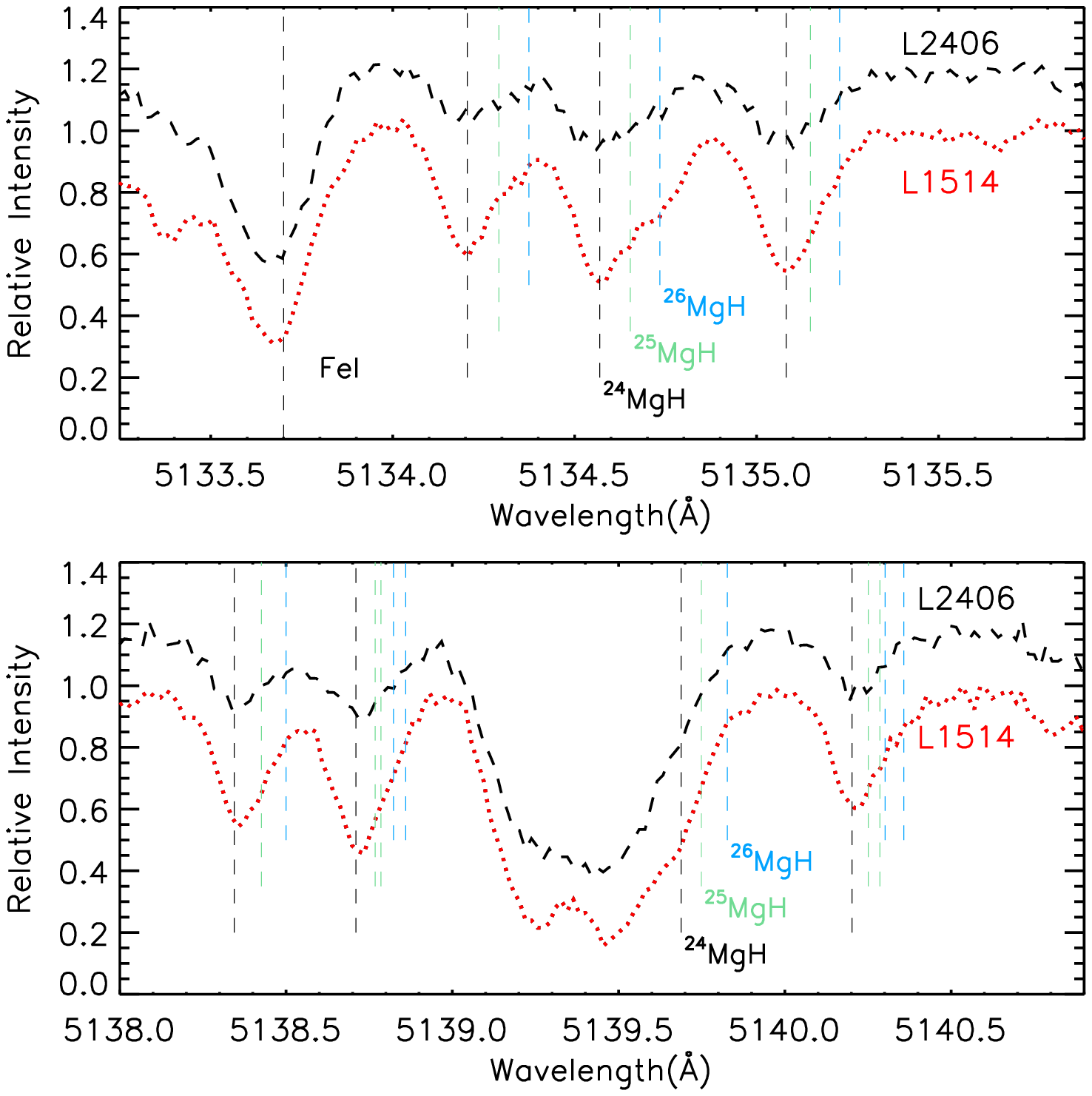}
\caption{The upper panel shows the wavelength region 5133\AA\/ to 5136\AA\/ and the lower panel 
shows the wavelength region 5138\AA\/ to 5141\AA\/ for the M4 primordial population star L1514 (red
dots) and for the M4 intermediate population star L2604 (black dashed line).  
As for Figs. \ref{DY1} and \ref{DY2} the positions of the $^{24}$MgH, $^{25}$MgH, and $^{26}$MgH 
lines are marked by vertical lines.   \label{DY3}
}

\end{figure}

Figures \ref{DY1} and \ref{DY2} show the final reduced spectra for the $\omega$~Cen stars over
wavelength regions that include electronic transitions of the MgH molecule, and the features specific 
to each Mg isotope are identified.  Similarly, Fig.\ \ref{DY3}
shows the isotopic ratio sensitive regions for the two M4 stars.  Apparent differences in the isotopic compositions are immediately evident from these figures, especially 
for the \wcen stars.  We now seek to quantify the differences. 
 
\section{ABUNDANCE ANALYSIS}

The isotopic abundances of magnesium were measured using the identical approach to that in 
\citet{DY03, DY04, DY06}  and thus can be briefly described here.  We used one dimensional local thermodynamic equilibrium (LTE) model atmospheres from \citet{kurucz96} and adopted stellar 
parameters for the $\omega$~Cen stars from \citet{ND95b}, from \citet{II99} for the M4 stars and from
\citet{DY03} for NGC 6752 mg0.  The particular stellar parameters employed
are not crucial as \citet{DY04} has shown that the derived isotope ratios appear insensitive to the
adopted stellar parameters, non-LTE effects and inadequacies in the model atmospheres.  
We analysed the three lines from the $A-X$ electronic transition of the MgH
molecule recommended by \cite{McL88} as providing the most reliable isotope
ratios (the numerous other MgH lines present in the spectra are more strongly
affected by known, and unknown, blends). The three lines are ($i$) 5134.6\AA\
from the $Q_1$(23) and $R_2$(11) lines of the (0,0) band, ($ii$) 5138.7\AA\
from the (0,0) $Q_1$(22) and (1,1) $Q_2$(14) lines, and ($iii$) 5140.2\AA\ from
the (0,0) $R_1$(10) and (1,1) $R_2$(4) lines. 

We note that other authors choose to analyse different MgH features -- for example, \citet{MC09}
prefer the 5134.3\AA\ feature to that at 5134.6\AA\ -- potentially resulting in systematically different
isotopic ratios.  We therefore make no strong claim for the validity of the absolute values of our
isotopic ratios but argue that consistent analysis allows meaningful comparisons in a relative sense.
It is also likely that known and unknown blends affecting the MgH analysis are more problematic for
cooler and more metal-rich stars, potentially leading to feature-to-feature differences in the derived 
isotopic ratios.  Non-LTE and 3D effects may also become more important for cooler temperatures. 
Some care must then be taken in comparing results as a function of 
metallicity although as shown below, our results are consistent with published values.  For example, 
our Mg isotopic abundance determination for the cool red giant HR 2140 
(T$_{eff}$ = 4000 K, log g = 0.9, [Fe/H] = --0.9 \citep{McL88}) is consistent with that of \citet{McL88}.  
Indeed the comparison of our results with published values that use the same MgH wavelength
regions argues that our results are internally consistent, although a relative offset cannot be ruled out. 
Nevertheless, additional studies seeking to identify blends affecting these and other MgH features would
be welcome.

Synthetic spectra were generated using the LTE stellar line analysis program
MOOG \citep{Sneden73}, with the line broadening, assumed to have a Gaussian form, 
estimated for each star by fitting the profile of the 5115.4\AA\ Ni\,{\sc i} line. The values of the line
broadening employed range from 6 to 10 km s$^{-1}$, with an uncertainty of order 
1--2 km s$^{-1}$.  
Each of the three MgH lines was analysed independently with three free parameters: the \mgb/\mga\/
and \mgc/\mga\/ isotopic ratios and the total Mg abundance, $\log \epsilon$(Mg).  A large range of
parameter space was explored and the optimum value for each parameter  identified by locating the 
$\chi^2$ minima.  Following \citet{Bev92}, the 1-$\sigma$ confidence limit for a given parameter 
was taken as the value at which $\Delta\chi^2$ = $\chi^2$ $-$ $\chi^2_{\rm min}$ = 1. 
Thus, for each line in each star, we obtain an optimum value and an uncertainty for \mgb/\mga\/ and  
\mgc/\mga.  The final ratio \mga:\mgb:\mgc\/ was then calculated as the weighted mean of the three 
determinations.  The individual determinations and the final adopted values for the \wcen 
and M4 stars, and NGC 6752 mg0, are given in Table \ref{Table 2}.

When expressing the isotopic ratios as \mga:\mgb:\mgc\/ =  (100 $-$ $b$ $-$ $c$):$b$:$c$, the formal 
errors in the determinations from each region are small; $b$ $\pm$ 1 and $c$ $\pm$ 1.  However,  
inspection of the values in Table \ref{Table 2}, for example, suggests that the real uncertainties are larger, 
being driven by systematic uncertainties in the continuum fitting, the adopted macroturbulent velocities,
and (known and unknown) blends.  As discussed in previous studies of stellar Mg isotope ratios,
the \mgb/\mga\/ ratio is often more uncertain than the \mgc/\mga\/ ratio since the isotopic shift is 
smaller, and as a result, the $^{25}$MgH feature is less well separated from the (generally) dominant  
$^{24}$MgH line. This effect can be seen qualitatively in what follows in Figures \ref{fig:43_100},
\ref{fig:132_150}, and \ref{fig:201_248} and quantitatively from the $\chi^2$
analysis.  We also note that the isotope ratios derived from each of
the three regions are usually, but not always, in good agreement. In some rare cases
(e.g., $\omega$~Cen ROA 201), the ratios from the three regions show poor agreement for which 
we have no obvious explanation.

An additional assessment of the uncertainties can be obtained from independent analysis of repeat 
observations, and from a comparison of isotopic ratios found here with the results of others.  For example,
three separate spectra of Arcturus, obtained during the AAT/UHRF runs, give \mga:\mgb:\mgc\/ ratios of 
81:9:10, 84:6:10 and 80:10:10 which are all in good agreement with the value 82:9:9 found by
\citet{McL88}.  There are also three separate AAT/UHRF spectra of the star
HR~2140.  As for Arcturus, our determinations for this star of 80:10:10, 82:8:10, and 83:7:10 are in
excellent accord with each other and with the value 88:6:6 given in \citet{McL88}.  Similarly, during the
Gemini-S/b-HROS run three field stars with previous Mg isotope ratio determinations were observed.  
The stars
are HD~26575 (HR~1299), HD~64606 and HD~103036 and we find isotopic ratios of 81:5:14, 89:2:9 
and 87:2:11, respectively.  The published isotopic ratio determinations for the first two are 72:18:10 
and 82:10:8, respectively,
from \citet{GL00}, while \citet{DY03} list 94:0:6 for HD~103036.  Our values are consistent with
the previous determinations with the possible exception of the \mgb/\mga\/ ratio for $\sim$solar
abundance star HD~26575 where the \citet{GL00} value\footnote{The value of \mgb/\mga\/ for 
HD~26575 (0.25) is not shown in the the \mgb/\mga\/ versus [Fe/H] diagram of \citet{GL00} 
(their Figure 8) as 
it lies outside the range depicted, notably higher than the \mgb/\mga\/ values for stars of comparable 
metallicity.} is notably higher than the value from our analysis.  We note, however, that our isotopic ratios 
for this star are consistent with the 80:10:10 determination given by \cite{McL88}.  Perhaps more
significantly, we find from our Gemini-S/b-HROS observations an isotope ratio of 62:9:29 for the
star NGC~6752 mg0, which is in excellent agreement with the value of 63:8:30 given in \citet{DY03} 
from VLT/UVES spectra.  The isotopic ratios derived from the three wavelength regions 
and the mean values for this star are also given in Table \ref{Table 2}.  
Overall, based on these comparisons, we estimate
that the uncertainties in our determinations are at the level $b$ $\pm$ 4 and $c$ $\pm$ 4 when 
writing the ratio as \mga:\mgb:\mgc\/ = (100 $-$ $b$ $-$ $c$):$b$:$c$. 

\section{RESULTS}

\begin{figure}
\centering
\includegraphics[angle=-90.,width=0.54\textwidth]{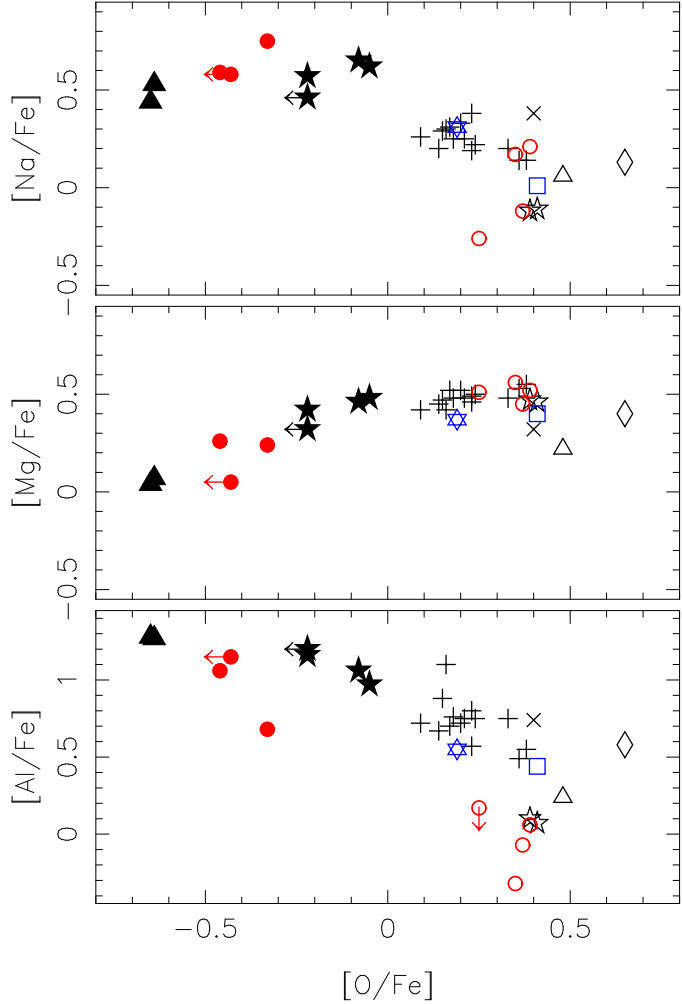} %results5_m4.plt
\caption{Sodium-to-iron (upper panel), magnesium-to-iron (middle panel), and aluminium-to-iron
(lower panel) abundance ratios are plotted against the oxygen-to-iron abundance ratio.
The three extreme population stars in $\omega$~Cen are shown as red filled
circles while the four primordial population red giants are plotted as red open circles.  The abundance
ratios are taken from \citet{ND95b}.  The M4 stars L1514 (primordial population, blue square) and
L2406 (intermediate population, blue six-pointed star) are plotted using the data of \citet{II99}.
The NGC~6752 data of \citet{DY03} are plotted as filled stars (extreme population), open 5-point 
stars (primordial population)
and plus-signs (intermediate population).  Similarly, using the data from \citet{DY06}, the two extreme 
population stars in M13 are plotted as filled triangles, the single primordial population star is shown as an
open triangle and the intermediate population star is plotted as an x-symbol.  The sole M71 star
analysed in \citet{DY06}, which is a primordial population star, is plotted as an open diamond.
 \label{namgal_o}}
\end{figure}

In the following we will make use of the results of \citet{DY03} for the red giants in NGC~6752 and of 
\cite{DY06} for M13 and M71, which have been derived using essentially the same procedure as that 
adopted here.
For NGC~6752, using the terminology of \citet{CB09} and the sodium and oxygen abundances 
given in \citet{DY03}, stars NGC~6752 mg12 and NGC~6752 mg24 are classified as members of the
primordial population, while stars NGC~6752 mg0, mg21, mg22 and 702 are members of the extreme
population.  The other 14 red giants are members of the intermediate population.  Similarly, for the four
M13 red giants studied in \citet{DY06}, and again using the sodium and oxygen abundances from
that work, star L598 is a member of the primordial population, star L629 is a member of the intermediate 
population and stars L70 and L973 both belong to the extreme population.   These 4 stars are a subset
of those studied by \citet{MS96}.  \citet{DY06} separate \mgb\/ from \mgc\/ but both studies find similar
(\mgb\/+\mgc\/)/\mga\/ ratios for a given star.  Further, the
single M71 star also analysed in \citet{DY06} is a CN-weak object \citep{MC09} with sodium and oxygen
abundances \citep{DY06, MC09} that classify it as a member of the primordial population of the cluster.
The sodium-to-iron, magnesium-to-iron and aluminium-to-iron abundance ratios as a function of [O/Fe]
for all these stars are shown in Fig.\ \ref{namgal_o}, along with the equivalent data for the \wcen 
and M4 stars from Table \ref{Table 1}.

\subsection{$\omega$ Cen ROA 94}

For this metal-poor star we find \mga:\mgb:\mgc\/ values of 88:2:9 with the standard uncertainty in the 
ratios of $b \pm 4$ and $c \pm 4$ where \mga:\mgb:\mgc\/ = $(100-b-c):b:c$.  The relative abundances 
of \mgb\/ and \mgc\/ found here for this star are somewhat higher, particularly for \mgc, 
than those for field halo dwarfs at this 
metallicity, which are typically 97:1:2 \citep[e.g.,][]{MC07}. 
However, the relative Mg isotope abundances appear to be somewhat lower than the 
$\sim$80:10:10 values
found for the primordial red giants in NGC~6752 and M13 \citep{DY03, DY06}.  The [Fe/H]
values for these clusters are higher, by $\sim$0.2 dex, than that of ROA~94.

\subsection{$\omega$ Cen ROA 43/100 and ROA 132/150}

\begin{figure}
\centering
\includegraphics[angle=0.,width=0.8\textwidth]{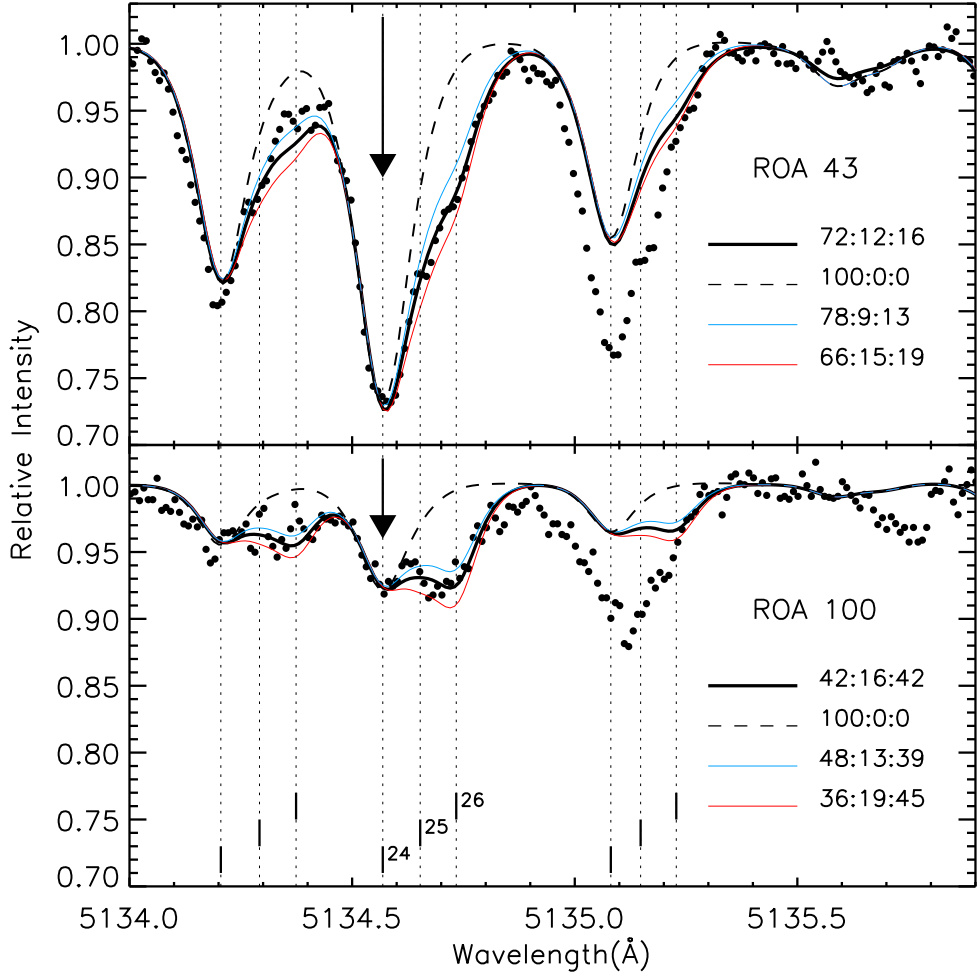} 
\caption{Spectra of $\omega$~Cen ROA 43 (upper panel) and ROA 100 (lower panel) compared with
synthetic spectra.  The observations are shown as filled circles and synthetic spectra with different
isotope ratios are overplotted.  The solid black line is the best fit, the dashed line
shows a pure \mga\/ mixture, and the thin colored lines show unsatisfactory ratios. The locations of the
$^{24}$MgH, $^{25}$MgH, and $^{26}$MgH lines are indicated.  The MgH feature
we are fitting is marked by the arrow.
 \label{fig:43_100}}
\end{figure}

\begin{figure}
\centering
\includegraphics[angle=0.,width=0.8\textwidth]{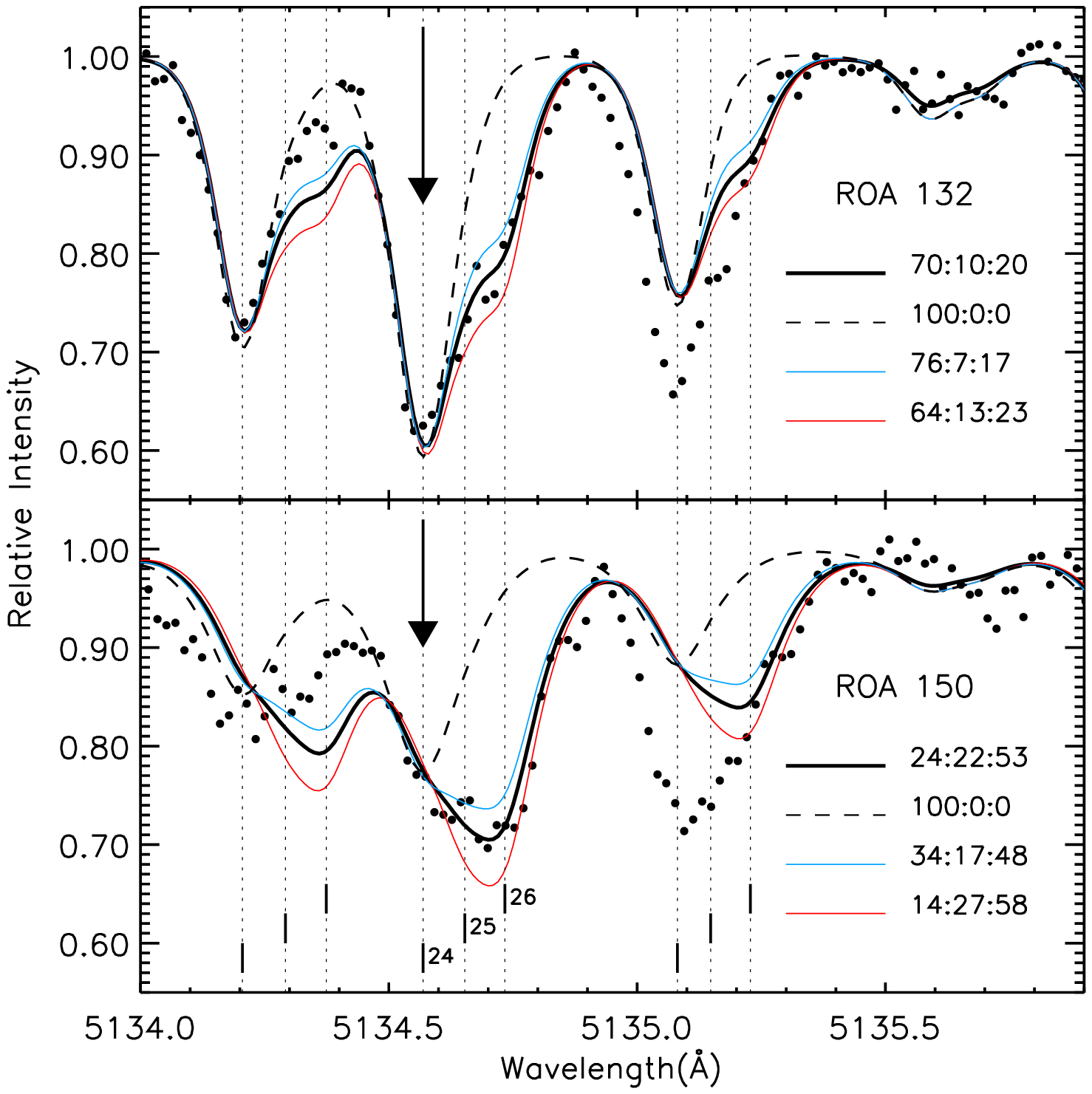} 
\caption{As for Figure \ref{fig:43_100} except for $\omega$~Cen stars ROA~132 (upper panel) and 
ROA~150 (lower panel).
 \label{fig:132_150}}
\end{figure}

As the derived Mg-isotope ratios for these two pairs of stars are found to be similar
(see Table \ref{Table 2}), we discuss them together.  Figures \ref{fig:43_100} and \ref{fig:132_150} 
show our fits to the 5134.6\AA\/ MgH feature; the fits to the 5138.7\AA\/ and 5140.2\AA\/ MgH
features are similar.  For the primordial stars ROA~43 and ROA~132 we find 
essentially identical relative abundances of \mga:\mgb:\mgc\/ = 76:10:14, with again uncertainties 
of order $\pm$4 in the $b$ and $c$ values.
These values are in reasonable agreement with the $\sim$80:10:10 values found
for the primordial stars in NGC~6752 and M13 by \citet{DY03, DY06}.  This is perhaps not surprising
given that the [Fe/H] values for NGC~6752 and M13 are only $\sim$0.1 to 0.2 dex lower than the
 [Fe/H] values of the $\omega$~Cen stars, which are given in Table \ref{Table 1}.

For the two extreme population stars we find \mga:\mgb:\mgc\/ = 48:13:39 for  ROA~100 and 
\mga:\mgb:\mgc\/ = 54:17:30 for ROA~150.  Here, given the larger contribution to the features from
the heavier isotopes and the lower overall Mg abundance (see Table \ref{Table 1}), the uncertainties 
in the $b$ and $c$ values are somewhat larger, of order $\pm 10$.   Nevertheless, the requirement for 
significant contributions from the heavier isotopes to permit a good fit to the observations is readily
apparent in the figures.  The isotope ratios for these 
$\omega$~Cen extreme population stars are also similar to those found for the extreme populations
in the clusters NGC~6752 and M13 \citep{DY03, DY06}, namely 62:7:31 for NGC~6752 (average of
4 stars) and 51:14:35 for M13 (average of 2 stars).  Again the iron abundances for these two clusters 
are only $\sim$0.1--0.3 dex lower than those of ROA~100 and ROA~150 (see Table \ref{Table 1}).

In both pairs the primary difference between the primordial and extreme stars is a
decrease in the \mga\/ relative abundance and an increase in the \mgc\/ relative abundance; the 
increase in the \mgb\/ relative abundance appears more modest.  We note also that the difference
in total magnesium abundance between the primordial and extreme stars in each pair, derived as part 
of the fitting process, is consistent at the 0.1--0.15 dex level with the differences inferred from 
Table \ref{Table 1}.  As argued by \citet{DY04} the total Mg abundances derived from the
fits to the MgH features are not as reliable as those determined from individual Mg\,{\sc i} lines;
nevertheless, the overall agreement is reassuring.
%{\color{red} For info, $\Delta$Mg for 43/100 is 0.47 from Table 1 and 0.59 from the MgH fits.  For 
%132/150, the corresponding numbers are 0.30 and 0.15, so these pairs are consistent.  For 201/248 
%things aren't as rosy - $\Delta$Mg is 0.21 from Table 1 but --0.12 from MgH.}

\subsection{$\omega$ Cen ROA 201/248}

\begin{figure}
\centering
\includegraphics[angle=0.,width=0.8\textwidth]{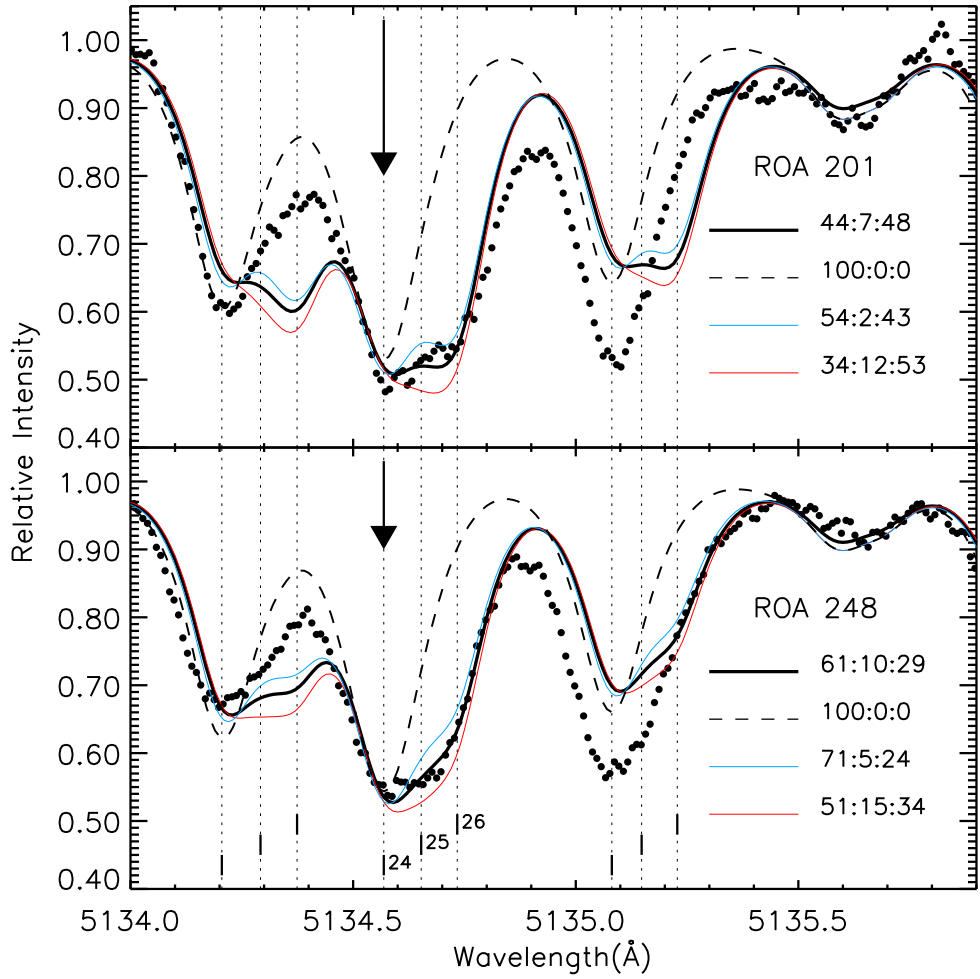} 
\caption{As for Figure \ref{fig:43_100} except for $\omega$~Cen stars ROA~201 (upper panel) and 
ROA~248 (lower panel).
 \label{fig:201_248}}
\end{figure}

For the O-rich, Na-poor primordial population star $\omega$~Cen ROA~201 ([Fe/H] = --0.85) 
our analysis yields 
\mga:\mgb:\mgc\/ = 74:19:7 while for the O-poor, Na-rich extreme population star $\omega$~Cen
ROA~248 ([Fe/H] = --0.78) the corresponding values are 77:10:13.  
Given that the uncertainties in the derived ratios 
are somewhat larger for these cool relatively metal-rich stars, it is not clear
whether we are seeing any significant difference in isotopic ratios.
%{\color{red} Unless you guys want it mentioned I'm not going to mention here that ND95 finds
%248 has a lower Mg abundance than 201 by 0.21 dex while the MgH analysis finds 248 has a higher 
% Mg abundance than 201 by 0.12 dex, because, as DY notes, the Mg abundances from the MgH fits 
% are quite sensitive to Teff.}
The synthetic spectrum fits to the 5134.6\AA\/ MgH feature in the observed spectra are 
shown in Fig.\ \ref{fig:201_248}.

The lack of obvious difference in Mg isotopic composition between these two $\omega$~Cen stars 
is consistent with the results of
\citet{MC09} who found only a small difference in Mg isotopic ratios between the CN-weak (primordial) 
and CN-strong (intermediate) populations in the relatively metal-rich cluster M71.  For example, the
\mgc/\mga\/ isotopic ratio for the CN-strong stars exceeds that for the CN-weak stars by only $\sim$4
percent \citep{MC09}.
The two M71 populations in their sample also do not show
any significant difference as regards total Mg abundance, and the ranges in O, Na and Al abundances
are small although they do exhibit the expected O-Na anti-correlation and Na-Al correlation
\citep{MC09}.   The [Fe/H] value for M71 \citep[--0.91,][]{DY06}, which we adopt here, is very similar 
to those for the $\omega$~Cen stars.  \citet{MC09} give [Fe/H] = --0.80 for M71.

\subsection{M4 L1514/L2406}

\begin{figure}
\centering
\includegraphics[angle=0.,width=0.8\textwidth]{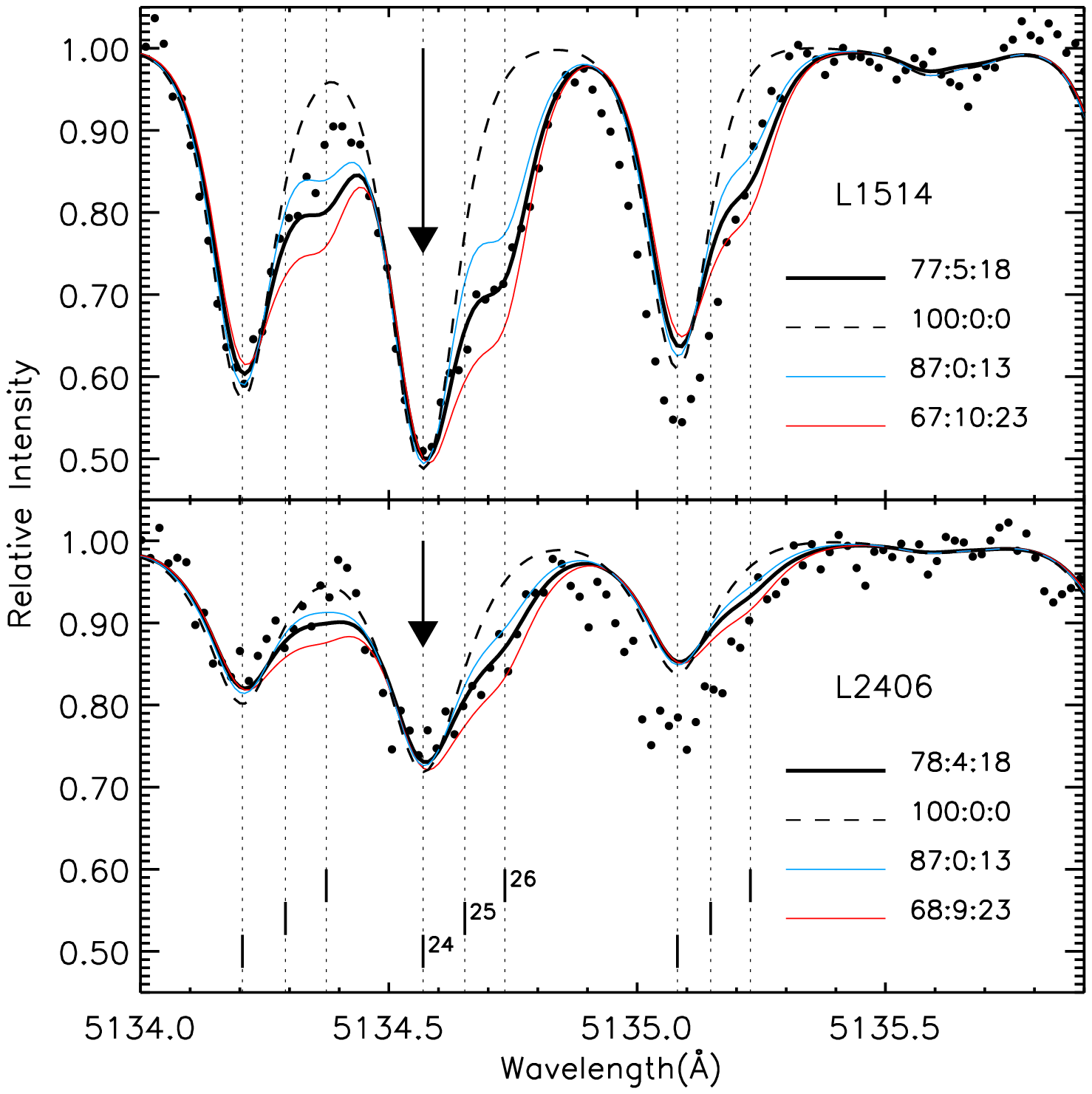} 
\caption{As for Figure \ref{fig:43_100} except for M4 stars L1514 (upper panel) and 
L2406 (lower panel).
 \label{fig:1514_2406}}
\end{figure}

Figure \ref{fig:1514_2406} shows the spectrum synthesis fits to the 5134.6\AA\/ MgH feature for
the two M4 stars.  The fits for the three wavelength regions yield mean isotopic ratios of 
82:6:12 for the primordial star L1514 and 80:6:15 for the intermediate population star L2406.  Given the
uncertainties in the determinations there is no apparent difference in the derived Mg isotope ratios.
\cite{DY08b} derived preliminary Mg isotope ratios for a number of M4 red giants.
Because their spectra had a resolution of $\sim$60,000, they assumed \citep[as did][]{MS96} that
\mgb\/ = \mgc.  For L1514 they found an isotope ratio of 80:10:10, which is consistent with our result.
An isotope ratio was not derived for L2406 but overall \citet{DY08b} did not find any significant 
evidence for a range in the Mg isotope ratios of their M4 stars.

The lack of any significant Mg isotope ratio difference for the two M4 stars observed here is perhaps 
not surprising given the lack of any difference in [Mg/Fe] and the (at most) small difference in [Al/Fe]  
for these stars \citep{II99,DY08}.  
Again this is consistent with the emerging general picture that while the O-Na anti-correlation 
is present at all metallicities, the extent of Al and especially Mg variations are much reduced at higher 
metallicities \citep[e.g.,][]{ND95b, II99, AM08,DY08b,MC09,JP10,DM11,Ca12}.

\section{DISCUSSION}

As discussed in \citet[][see also \citet{WW95,TWW95}]{KKU11}, the heavy isotopes of magnesium 
have their origin in a number of 
nucleosynthetic processes.  For example, explosive nucleosynthesis in massive star supernovae
can produce \mgb\/ and \mgc\/ through the He-burning reactions $^{22}$Ne($\alpha$, $n$)$^{25}$Mg
and $^{22}$Ne($\alpha$, $\gamma$)$^{26}$Mg.  Being ``secondary'' isotopes, the production of the 
neutron-rich isotopes increases with the metallicity of the supernovae, although some primary
production also occurs \citep{KKU11}.  
The nucleosynthetic model predictions \citep{KKU11} are at least qualitatively consistent with the 
observations of the variation of the Mg isotopic abundance ratios with [Fe/H] in the field star samples 
of \citet{DY03b} and \citet{MC07}.  

Relatively massive (M $>$ 4 M$_{\sun}$) AGB stars can also contribute
to the production of the heavy magnesium isotopes \citep[e.g.,][]{KL03, AK10, KKU11}.
In such stars there are a number of production sites: (1) the vicinity of the He-burning shell 
via $\alpha$-capture 
reactions on $^{22}$Ne; (2) via the Mg-Al cycle in the vicinity of the H-burning shell; and (3) 
at the base of the 
convective envelope during the hot-bottom burning phase, again via the Mg-Al cycle \citep{KL03}.  In all
cases the processed material can be brought to the surface layers and thus can contribute to the
enrichment of the surrounding interstellar medium via mass loss in a stellar wind.  As discussed in 
\citet{KL03} and \citet{,AK10} the yields of \mgb\/ and \mgc\/ are complex functions of the initial 
abundance of the
AGB star and of the parameters that control the thermal pulse structure, the extent of thermal dredge-up,
the conditions in the convective envelope and the rate of mass loss.   
In the context of the current work it is worth noting that 
the production of \mgb\/ and \mgc\/ via $\alpha$-capture reactions on $^{22}$Ne is unrelated to the 
abundance of aluminium.  
Consequently, since we find that the relative \mgc\/ abundance is tightly coupled 
to that of Al, at least for stars with [Al/Fe] $\gtrsim$ 0.5 (see top panel of Fig.\ \ref{iso_alfe}), 
we can conclude that it is unlikely that the shell He-burning production site is a significant contributor.
Indeed it is likely that the primary nucleosynthesis site is hot-bottom burning in relatively 
massive AGB stars.

\begin{figure}
\centering
\includegraphics[angle=-90.,width=0.6\textwidth]{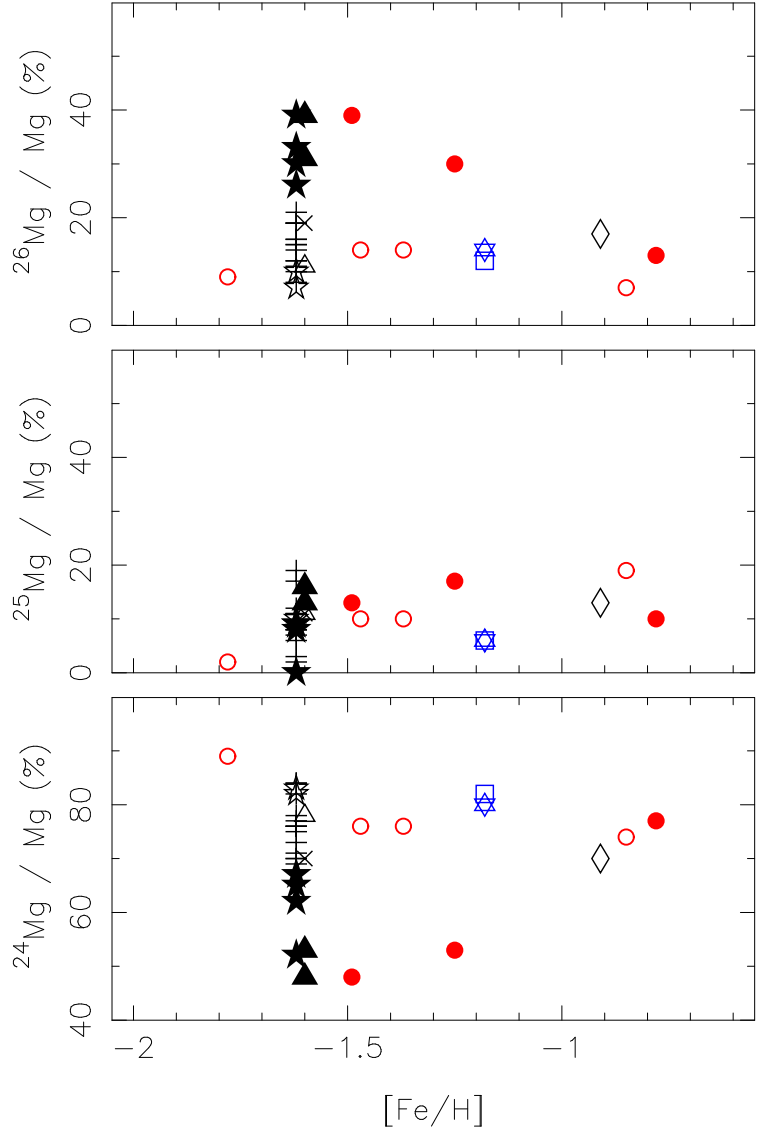} %results2_new2_m4.plt
\caption{Relative abundance fractions for \mga\/ (lower panel), \mgb\/ (middle panel) and \mgc\/ (upper
 panel) compared to the total Mg abundance plotted against
iron abundance [Fe/H].  Symbols are as for Fig.\ \ref{namgal_o}.
 \label{iso_feh}}
\end{figure}

\begin{figure}
\centering
\includegraphics[angle=-90.,width=0.6\textwidth]{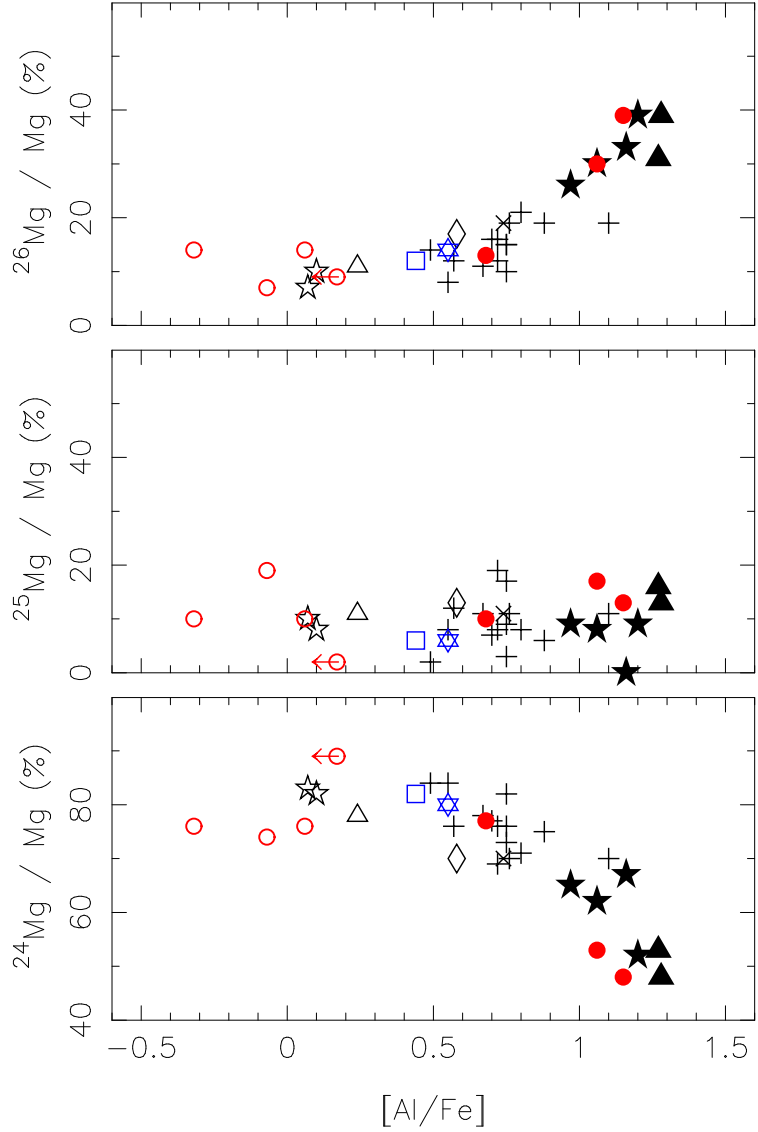} %results3_new2_m4.plt
\caption{Relative abundance fractions for \mga\/ (lower panel), \mgb\/ (middle panel) 
and \mgc\/ (upper panel) compared to the total Mg abundance plotted against [Al/Fe], the 
aluminium to iron abundance ratio.
Symbols are as for Fig.\ \ref{namgal_o}.
 \label{iso_alfe}}
\end{figure}

In the panels of Fig.\ \ref{iso_feh} we plot the individual isotopic fractions relative to the total Mg
abundance against [Fe/H] for the stars observed in $\omega$~Cen, using the \citet{ND95b}
[Fe/H] values, and for the M4 stars using the cluster mean [Fe/H] from \citet{II99}.  Also shown are the red 
giants observed in the globular clusters NGC~6752 \citep{DY03} and M13 and M71 \citep{DY06}.  
Since the observations in \citet{DY03, DY06} were
analysed using very similar techniques to those employed here, no systematic effects are expected and
the data can be combined\footnote{We assume that the [Fe/H] values are also free of systematic
effects.  This is supported to some extent by the fact the mean iron abundance found for NGC~6752 in
\citet{DY03}, [Fe/H] = --1.62, is in reasonable accord with the mean iron abundance, [Fe/H] = --1.52,
of the six NGC~6752  red giants analysed in \citet{ND95b}. Similarly, the [Fe/H] determinations for M4 in
\citet{II99} and \citet{DY08} are in excellent agreement.}.   
In the figure, filled symbols represent extreme population stars while open symbols represent primordial 
populations stars.  Fig.\ \ref{iso_alfe} shows the same relative abundance fractions
but in this case plotted against the [Al/Fe] abundance ratio.

Considering first the primordial stars in the middle and upper panels of Fig.\ \ref{iso_feh}, there does
not seem to be any obvious trend of increasing \mgb\/ and \mgc\/ fraction relative to total Mg with 
increasing [Fe/H], although it may be that any (small) trend is disguised by the errors in the determinations.
Indeed the location of the primordial-star points in the lower panel of figure, where effectively 
the \mgb\/ and \mgc\/ determinations are averaged, does support the existence of such a trend:
making no attempt to distinguish \mgb\/ from \mgc, the relative isotopic ratios appear to change smoothly
from $\sim$84:7:7 at [Fe/H] $\approx$ --1.8 to $\sim$70:15:15 at [Fe/H] $\approx$ --0.7 dex.  
Such a trend is qualitatively consistent with the nucleosynthetic predictions of 
increasing \mgb/\mga\/ and \mgc/\mga\/ with [Fe/H] \citep[e.g.,][]{KKU11}, although as noted in 
\citet{DY03, DY06}, for example, the isotopic ratios for these primordial
globular cluster stars are significantly higher 
than for field stars of comparable [Fe/H] abundance, despite being otherwise chemically similar.
\citet{DY03, DY06} postulate the existence of a generation of relatively massive AGB stars in
the proto-cluster environment.  These stars uniformly enrich the gas, from which the current
primordial population stars form, in the heavy isotopes of Mg to a level above that generated 
by metal-poor supernovae, which is represented by the Mg isotopic ratios for the general halo field 
population at that metallicity.   Given the similarity of our results for the \wcen and M4 primordial 
stars to those of 
\citet{DY03, DY06}, it is apparent that this postulated scenario could also apply to the 
proto-$\omega$ Cen and the proto-M4.

Turning now to the full sample of stars, it is noteworthy from the middle panels of Figs.\ \ref{iso_feh} 
and \ref{iso_alfe} that there does not seem to be any obvious difference between the primordial and the 
extreme stars
as regards the \mgb\/ fractional abundance, regardless of [Fe/H] and regardless of the [Al/Fe] ratio.  
The lack of correlation between \mgb\/ fractional abundance and [Al/Fe] has already been noted by
\citet{DY03,DY06} and it is apparent that the result applies to the \wcen 
and M4 stars observed here as well. 
Further, the intermediate population stars (predominantly from NGC~6752) also have similar values, and
a single \mgb\/ to total Mg fraction of order 10\% appears consistent with all the determinations.  Indeed
for the 34 stars shown in the figures, the mean \mgb\/ to total Mg fraction is 9.8\% with a standard
deviation of 4.6\%, which is comparable to the uncertainties in the individual determinations. 
In other words, in these globular cluster stars, which cover approximately 1 dex in [Fe/H] and 1.3 dex
in [Al/Fe],  and with the current uncertainties, the \mgb\/ relative abundance appears not to
be influenced by whether the star has been formed from the original cluster material, i.e.\ the primordial
population, or from significantly ``polluted'' material as in the extreme population stars, which are
presumed to incorporate material that has been processed through the Mg-Al cycle.   This lack of any 
obvious difference between the different populations is surprising
as, a priori, one would not expect the relative \mgb\/ abundance in Mg-Al cycle processed material to be 
essentially identical to that for unprocessed material.  In this context however, it is worth noting that 
\citet{MC09} do find a small difference in the \mgb\/ relative abundance between the CN-weak 
(primordial) and the CN-strong (intermediate) stars in the metal-rich globular cluster M71.  Specifically,
for 5 CN-weak stars, the mean value of the \mgb\/ to total Mg relative abundance is 5.2 $\pm$ 0.4
percent, while for 4 CN-strong stars the mean \mgb\/ relative abundance is slightly higher, 7.3 $\pm$ 0.4 
percent.  In both cases the error given is the standard error of the mean.

The situation for the \mgc\/ fractional abundance, however, is markedly different.  The upper panels
(and, given the lack of variation in \mgb, the effectively mirror images in the lower panels) of 
Figs.\ \ref{iso_feh} and \ref{iso_alfe} reveal clear dependencies on both [Fe/H] and on [Al/Fe].
In particular, Fig.\ \ref{iso_feh} shows that the production of \mgc\/ is notably less at higher overall 
abundance.
This is consistent with the fact that, in general, the extreme and intermediate population stars in 
globular clusters with larger [Fe/H] values show lower enhancements in [Al/Fe] compared to more
metal-poor systems \citep[e.g.,][]{ND95b, DY08,MC09,JP10, DM11,Ca12}.  This is usually interpreted
\citep[e.g.,][]{JP10} as a consequence of lower temperatures at the bottom of the convective
envelope in the more metal-rich relatively massive AGB stars that are assumed to be the source of the
[Al/Fe] enhancements.  Indeed, at least as regards $\omega$~Cen,
\citet{DA11} claim that the smaller amount of [Al/Fe] enhancement seen at higher [Fe/H] values is
consistent with the predictions of the \citet{VDA09} AGB star models.  The reduced production of
\mgc\/ at larger [Fe/H] seen in Fig.\ \ref{iso_feh} is then presumed to be consistent with the same 
interpretation, given the strong connection between the \mgc\/ fractional abundance and [Al/Fe] seen 
for the extreme stars in Fig.\ \ref{iso_alfe}.

As regards the data presented in the upper panel of Fig.\ \ref{iso_alfe}, and the implicit mirror image in the 
lower panel given the lack of variation in \mgb, it appears that there are effectively two
regions in the relation between \mgc\/ and [Al/Fe].  The first is occupied by the primordial and
intermediate population stars.  Here the [Al/Fe] ratio increases by approximately 0.6 dex from the low
value in the primordial stars but the \mgc\/ fraction remains essentially constant, as does the 
\mga\/ fraction.  The second region is occupied primarily by the extreme stars though it includes some
intermediate populations stars.  Here we see that the
\mgc\/ abundance fraction is tightly coupled with that of $^{27}$Al as the [Al/Fe] ratio rises from
$\sim$0.6 dex to the value of $\sim$1.2--1.3 in the most extreme population stars.  At the same time
the \mgc\/ fraction rises from approximately 10\% to almost 40\% while the \mga\/ fraction drops to 
$\sim$50\%.      

The first region of the relation can be understood if it is assumed 
that the increased Al abundance is generated from \mga\/ via the Mg-Al cycle.  For example, if we 
assume the following parameters: [Fe/H] = --1.62, [Mg/Fe] = +0.5, [Al/Fe] = 0.0 and Mg isotopic
ratios of 80:10:10 (i.e., those of the NGC 6752 primordial stars), then, ignoring any dilution, converting
just 10\% of the \mga\/ atoms present to $^{27}$Al results in an increase in [Al/Fe] of 0.7 dex while the
(total) [Mg/Fe] is reduced by only 0.05 dex.  Such a small change in [Mg/Fe] is within the current limits on
the variation in [Mg/Fe] for the NGC~6752 primordial and intermediate population stars \citep{DY03}.
Further, assuming that the abundance of the more massive Mg isotopes remains relatively unaltered, 
then the isotopic ratios change only to 78:11:11.  Hence in this region of the diagram,
the increase in [Al/Fe] and the lack of any significant change in the isotope ratios is 
consistent with the conversion of a small amount of \mga\/ to Al in the Mg-Al cycle.  We note that the data
shown in Fig.\ \ref{iso_alfe} appear to require the consumption to be primarily of \mga; 
if the synthesised Al 
atoms in our example come entirely from \mgb\/ or \mgc\/ then that would produce significant changes in
the \mgb\/ or \mgc\/ fractional abundances inconsistent with the observations shown in Fig.\ \ref{iso_alfe}.

In the second region of Fig.\ \ref{iso_alfe} the production of Al and \mgc\/ are apparently closely related.
This is somewhat surprising, as in the conventional treatment of the Mg-Al cycle, \mgc\/ is destroyed 
by proton captures (to form $^{27}$Al) and is not expected to be made in any significant amount.  
Specifically, the proton-capture reaction on \mgb\/ produces $^{26}$Al, which, in the ground-state, 
has a long \mbox{$\beta^{+}$-decay} time of $\sim$0.72 Myr.  Consequently, the cycle progresses 
from $^{26}$Al primarily through the pathway $^{26}$Al($p$,$\gamma$)$^{27}$Si($\beta^{+}$)$^{27}$Al
and no \mgc\/ is synthesised. $^{26}$Al does, however, possess an excited state isomer, 
$^{26}$Al$^{m}$, for which the $\beta^{+}$-decay time is short (6.3 sec, comparable to the 
$\beta^{+}$-decay time of $^{27}$Si) thus providing a channel to produce \mgc.  Unfortunately, 
\citet{IC11} indicate that the rate for the reaction $^{25}$Mg($p$,$\gamma$)$^{26}$Al$^{m}$ is 
approximately a factor of five lower than the rate for the equivalent reaction involving the ground-state 
of $^{26}$Al at the temperatures likely to apply in the hot-bottom burning regions in relatively massive 
AGB stars.  Consequently, the generation of $^{27}$Al via the Mg--Al cycle is not expected
to show the strong correlation between $^{27}$Al and \mgc\/ seen in Fig.\ \ref{iso_alfe}.

In the context of the heuristic example outlined above, if we now suppose 50\% (rather than 10\%) of the 
available Mg atoms are processed, including half of the original \mgb\/ (but not \mgc\/), 
and that the output of the processing is 80\% $^{27}$Al and 20\% \mgc\/ (or equivalently 
20\% $^{26}$Al that
is assumed to ultimately decay to \mgc) then the resulting product has [Mg/Fe] = 0.28, 
[Al/Fe] = 1.24 and \mga, \mgb, and \mgc\/ fractions of the total Mg abundance of 58\%, 8\% and 33\%,
respectively.  These values are consistent with the data of Fig.\ \ref{iso_alfe} for the extreme stars.
Moreover, the change in [Mg/Fe] from +0.5 to +0.28 is consistent with the lower [Mg/Fe] abundance
ratios seen in the NGC~6752 extreme population stars \citep[][see also Table \ref{Table 1}]{DY03}.
While these calculations are admittedly ad hoc, they do serve to illustrate that 
in order to understand the (\mgc, [Al/Fe]) relation revealed by the extreme population stars in 
Fig.\ \ref{iso_alfe}, synthesis of both $^{27}$Al {\it and} \mgc\/ from \mga\/ would seem to be required.

The discrepancy between the observational results presented in Fig.\ \ref{iso_alfe} and
current theory is highlighted by a comparison with the AGB-star evolutionary models
presented in \citet{V11}.  Their ``reference model''  is for a 6 M$_{\sun}$ star with a composition
typical of intermediate metallicity globular clusters ([$\alpha$/Fe] = +0.4 and Z = 0.001 or 
[M/H] $\approx$ --1.3) evolved through the complete AGB phase \citep[see][for details]{V11}.  
The temperature at the
bottom of the convective envelope reaches sufficiently high values that the Mg-Al cycle is initiated
and over the course of the AGB evolution, i.e., prior to the loss of most of the envelope via mass loss, 
in the model the
surface Mg abundance declines by about 0.35 dex while the surface Al abundance rises by $\sim$1 dex,
which is in approximate agreement with the observations.
However, in this model \mgc\/  is destroyed throughout the AGB evolution by the $p$-capture reaction
and not synthesised, so that there is little real change, from its initial low value, in the \mgc\/  fraction of the 
total surface Mg abundance (which decreases during the evolution).  
In contrast, with the reaction rates and the treatment of thermal pulses, 3rd dredge-up and convective
mixing used in the model \citep[see][for details]{V11}, the total surface Mg 
abundance becomes dominated by that of \mgb\/, so that by the end of the evolution the \mgb\/
fraction of the total Mg abundance exceeds 90\% \citep{V11}.  It is very difficult to see how these
surface fractional abundances can be reconciled with the observations of the extreme population
stars shown in Fig.\ \ref{iso_alfe}, even allowing for dilution of the AGB star surface abundances with
pristine material, which is not depleted in overall Mg abundance and which has fractional isotopic
abundances like those of the primordial stars.  

We speculate here on two possibilities that might give rise to a higher \mgc\/ fractional abundance
in the most Al-rich stars assuming, given the strong correlation seen in Fig.\ \ref{iso_alfe}, that the
general context is that of the Mg-Al cycle and AGB star evolutionary calculations such as 
those of \citet{V11}.  In the first scenario we postulate that the \mgc\/ is synthesised as \mga\/ is depleted.  
To achieve this we suggest, despite the results of \citet{IC11}, an 
increase in the rate for the reaction $^{25}$Mg($p$,$\gamma$)$^{26}$Al$^{m}$ that produces 
the excited state isomer $^{26}$Al$^{m}$.  As noted above this isomer rapidly $\beta^{+}$-decays
to produce \mgc\/.
An increase in this rate and in that for 
$^{25}$Mg($p$,$\gamma$)$^{26}$Al$^{g}$ would also decrease the \mgb\/ fractional abundance
reducing the ``over-production'' of \mgb\/ seen in the \citet{V11} models.  In this scenario the
rate for the reaction $^{26}$Mg($p$,$\gamma$)$^{27}$Al may also have to be reduced to ensure
the synthesised \mgc\/ is not entirely processed further to $^{27}$Al. 

In the second scenario we postulate that $^{26}$Al (as distinct from \mgc) is synthesised as \mga\/ 
is depleted.  Here we note that once Mg-Al cycle processing
has ceased, material rich in $^{26}$Al will eventually show an increase in the \mgc\/ relative 
abundance as the $^{26}$Al radioactively decays to \mgc.  To achieve this we suggest a significant 
increase in the $^{25}$Mg($p$,$\gamma$)$^{26}$Al$^{g}$
reaction rate (and probably also a reduction in the rate of the $^{26}$Al($p$,$\gamma$)$^{27}$Si 
reaction, although $^{27}$Al still needs to be produced from the $\beta^{+}$-decay of $^{27}$Si).
Here, again in the context of the \citet{V11} models, the destruction of \mga\/ does not 
primarily produce \mgb\/ but rather instead results in significant amounts of $^{26}$Al (as well as 
$^{27}$Al).  This ``over-produced'' $^{26}$Al is then subsequently 
observed as \mgc\/, the radioactive decay product of $^{26}$Al.  The suggested rate increase
would also serve to alleviate the discrepancy between the theoretical predictions and the observations
as regards the \mgb\/ relative abundance.  

It is worth noting that the exploratory H-burning nucleosynthesis calculations of \citet{PCI07}, which 
were carried 
out at a variety of {\it fixed temperatures rather than in the context of a full evolutionary stellar model}, 
and which employed the best available reaction rate data, were able to reproduce the 
Mg isotope results for the globular cluster NGC~6752 given in \citet{DY03} as well as the distributions
of the O, Na and Al abundances.  The agreement required quite 
specific conditions in their relatively simple analysis: a temperature of approximately 75MK (below 70MK 
\mga\/ is hardly affected while above 80MK it is rapidly destroyed) and for the extreme stars, a dilution
of the processed material via the incorporation of approximately 30\% material having the primordial 
composition.  They also point out that the ``excess'' of \mgc\/ seen in the NGC~6752 stars can only be 
understood if it is originally produced as $^{26}$Al \citep{PCI07}.
Whether these fixed temperature results apply to the more complex situation in real stars is a question 
that is beyond the scope of this paper.  

We note only that our results for the Mg isotopic ratios in
\wcen extreme population stars reinforce the discrepancy between 
observations and stellar evolutionary theory already apparent in the results of \citet{MS96}
and \citet{DY03, DY06}.  
A resolution of the discrepancy is likely to lead to further insight into the complicated nucleosynthetic 
processes that occurred early in the lifetime of most globular clusters and in that respect further 
determinations of magnesium isotopic ratios in additional globular cluster stars would be very worthwhile. 
The data presented in Figs.\ \ref{iso_feh} and \ref{iso_alfe} suggests strongly that the stars to target in
such a program, those most likely to have significantly non-solar Mg isotopic ratios, are the stars with 
the highest enhancements of [Al/Fe] in globular clusters with intermediate or lower metallicities, 
i.e.\ [Fe/H] $\lesssim$ --1.4 dex. 

\acknowledgments

GSDC and JEN would like to acknowledge the provision of travel funds for the Gemini-S observing run 
from the Australian Government Access to Major Research Facilities Program, Grant 05/06-O-15.
JEN and DY are also pleased to acknowledge support from Australian Research Council Discovery
Projects grant DP0984924.
The authors are grateful to Inese Ivans for carrying out some of the AAT observations and to the
then Gemini-South Director Michael West for authorising a
classical-night to queue-night swap when instrument problems looked likely to cause the complete 
loss of our second scheduled Gemini-S night.  Thanks also to the Gemini staff for fixing the problems 
allowing completion of our program on the following night.  The support of b-HROS scientist
Steve Margheim during the run is also appreciated, as was the assistance of Sean Ryan during the 
original investigation of the suitability of UHRF for this observational program.  Useful conversations 
with Amanda Karakas and John Lattanzio are also gratefully acknowledged.

This work is based in part on observations 
for program GS-2006A-Q-44 obtained at the Gemini Observatory, which is operated by the 
Association of Universities for Research in Astronomy, Inc., under a cooperative agreement 
with the NSF on behalf of the Gemini partnership: the National Science Foundation (United 
States), the Science and Technology Facilities Council (United Kingdom), the 
National Research Council (Canada), CONICYT (Chile), the Australian Research Council (Australia), 
Minist\'{e}rio da Ci\^{e}ncia, Tecnologia e Inova\c{c}\~{a}o (Brazil) 
and Ministerio de Ciencia, Tecnolog\'{\i}a e Innovaci\'{o}n Productiva (Argentina).

{\it Facilities:} \facility{AAT}, \facility{Gemini-South}

\begin{deluxetable}{lcccccc}
\tablewidth{0pt}
\tablecaption{Globular Cluster Red Giant Mg isotope ratios \label{Table 2}}
\tablecolumns{7}
\tablehead{
\colhead{Star} & 
\colhead{Region} & 
\colhead{$^{25}$Mg/$^{24}$Mg} & 
\colhead{$\sigma$} & 
\colhead{$^{26}$Mg/$^{24}$Mg} & 
\colhead{$\sigma$} & 
\colhead{$^{24}$Mg:$^{25}$Mg:$^{26}$Mg} }
\startdata
$\omega$ Cen ROA 94            & 5134.6\AA\ & 0.000 & 0.031 & 0.190 & 0.023 & 84:00:16 \\
                 & 5138.7\AA\ & 0.006 & 0.054 & 0.029 & 0.038 & 97:01:03 \\
                 & 5140.2\AA\ & 0.139 & 0.091 & 0.000 & 0.066 & 88:12:00 \\
                 & $Mean$       & 0.027 & 0.016 & 0.106 & 0.012 & 88:02:09 \\
                                                                         \\
$\omega$ Cen ROA 43            & 5134.6\AA\ & 0.161 & 0.015 & 0.225 & 0.010 & 72:12:16 \\
                 & 5138.7\AA\ & 0.167 & 0.021 & 0.148 & 0.011 & 76:13:11 \\
                 & 5140.2\AA\ & 0.066 & 0.023 & 0.194 & 0.016 & 79:05:15 \\
                 & $Mean$       & 0.137 & 0.006 & 0.190 & 0.004 & 75:10:14 \\
                                                                         \\
$\omega$ Cen ROA 100           & 5134.6\AA\ & 0.368 & 0.068 & 1.005 & 0.065 & 42:16:42 \\
                 & 5138.7\AA\ & 0.011 & 0.086 & 1.087 & 0.097 & 48:01:52 \\
                 & 5140.2\AA\ & 0.497 & 0.115 & 0.461 & 0.072 & 51:25:24 \\
                 & $Mean$       & 0.282 & 0.029 & 0.835 & 0.025 & 47:13:39 \\
                                                                         \\                                                                         
$\omega$ Cen ROA 132           & 5134.6\AA\ & 0.148 & 0.012 & 0.279 & 0.010 & 70:10:20 \\
                 & 5138.7\AA\ & 0.064 & 0.017 & 0.204 & 0.011 & 79:05:16 \\
                 & 5140.2\AA\ & 0.176 & 0.020 & 0.061 & 0.012 & 81:14:05 \\
                 & $Mean$       & 0.130 & 0.005 & 0.188 & 0.004 & 76:10:14 \\
                                                                         \\
$\omega$ Cen ROA 150           & 5134.6\AA\ & 0.908 & 0.091 & 2.189 & 0.106 & 24:22:53 \\
                 & 5138.7\AA\ & 0.124 & 0.042 & 0.408 & 0.031 & 65:08:27 \\
                 & 5140.2\AA\ & 0.203 & 0.053 & 0.234 & 0.029 & 70:14:16 \\
                 & $Mean$       & 0.312 & 0.019 & 0.550 & 0.013 & 54:17:30 \\
                                                                         \\
$\omega$ Cen ROA 201           & 5134.6\AA\ & 0.163 & 0.013 & 1.095 & 0.036 & 44:07:48 \\
                 & 5138.7\AA\ & 0.278 & 0.025 & 0.036 & 0.004 & 76:21:03 \\
                 & 5140.2\AA\ & 0.497 & 0.036 & 0.000 & 0.008 & 67:33:00 \\
                 & $Mean$       & 0.259 & 0.007 & 0.098 & 0.002 & 74:19:07 \\
                                                                         \\
                                                                         \tablebreak
$\omega$ Cen ROA 248           & 5134.6\AA\ & 0.165 & 0.014 & 0.481 & 0.017 & 61:10:29 \\
                 & 5138.7\AA\ & 0.078 & 0.009 & 0.082 & 0.004 & 86:07:07 \\
                 & 5140.2\AA\ & 0.241 & 0.027 & 0.230 & 0.012 & 68:16:16 \\
                 & $Mean$       & 0.134 & 0.005 & 0.173 & 0.003 & 77:10:13 \\
                 \\
M4            L1514            & 5134.6\AA\ & 0.058 & 0.007 & 0.226 & 0.008 & 78:05:18 \\
                 & 5138.7\AA\ & 0.111 & 0.014 & 0.062 & 0.010 & 85:09:05 \\
                 & 5140.2\AA\ & 0.085 & 0.013 & 0.109 & 0.014 & 84:07:09 \\
                 & $Mean$       & 0.078 & 0.003 & 0.142 & 0.003 & 82:06:12 \\
                                                                         \\
M4 L2406            & 5134.6\AA\ & 0.054 & 0.022 & 0.226 & 0.015 & 78:04:18 \\
                 & 5138.7\AA\ & 0.053 & 0.038 & 0.243 & 0.025 & 77:04:19 \\
                 & 5140.2\AA\ & 0.129 & 0.040 & 0.073 & 0.022 & 83:11:06 \\
                 & $Mean$       & 0.073 & 0.010 & 0.185 & 0.007 & 80:06:15 \\
                                                                         \\
NGC~6752 mg0        & 5134.6\AA\ & 0.163 & 0.019 & 0.495 & 0.014 & 60:10:30 \\
                 & 5138.7\AA\ & 0.029 & 0.027 & 0.535 & 0.019 & 64:02:34 \\
                 & 5140.2\AA\ & 0.285 & 0.038 & 0.339 & 0.023 & 62:18:21 \\
                 & $Mean$       & 0.148 & 0.009 & 0.467 & 0.006 & 62:09:29 \\
                                                                         \\
\enddata
\end{deluxetable}

\end{document}